\documentclass[twocolumn]{aastex631}
\usepackage{savesym}
\usepackage{tipa} 
\savesymbol{tablenum}
\usepackage{siunitx}
\restoresymbol{SIX}{tablenum}
\DeclareSIUnit{\jansky}{Jy}
\DeclareSIUnit{\MSPS}{MSPS}
\DeclareSIUnit{\byte}{B}
\DeclareSIUnit{\tecu}{TECu}
\DeclareSIUnit{\bit}{bit}
\DeclareSIUnit{\sample}{S}
\DeclareSIUnit{\dmunit}{pc~cm^{-3}}
\DeclareSIUnit{\millisec}{ms}
\newcommand{\kkoname}{k'ni\textipa{P}atn k'l$\left._\mathrm{\smile}\right.$stk'masqt}
\newcommand{\rmtec}{\mathrm{TEC}}

\usepackage{cancel}

\usepackage{aas_macros}

\usepackage{amsmath, amssymb,commath}
\usepackage{comment}
\usepackage{tikz}
\newcommand{\corrname}{\texttt{PyFX}}

\newcommand{\kdm}{K_{\textrm{DM}}}
\DeclareSIUnit{\parsec}{pc}

\newcommand{\HCOPOS}{(-2523644.2m,-4123700.4m,4147773.46m) }
\newcommand{\GBOPOS}{(883728.0m, -4924463.3m,  3943957.56m)  }

\newcommand{\ncaldump}{300 }
\newcommand{\npulsardumps}{200 }

\received{\today}
\revised{\today}
\submitjournal{ApJ}
\shorttitle{VLBI Astrometric Procedure and Performance with CHIME/FRB Outriggers}
\shortauthors{authors}

\begin{document}
\title{VLBI Astrometric Procedure and Performance with CHIME/FRB Outriggers}
\author[0000-0002-3980-815X]{Shion Andrew}
\affiliation{MIT Kavli Institute for Astrophysics and Space Research, Massachusetts Institute of Technology, 77 Massachusetts Avenue, Cambridge, MA 02139, USA}
\affiliation{Department of Physics, Massachusetts Institute of Technology, 77 Massachusetts Avenue, Cambridge, MA 02139, USA}

\author[0000-0002-5857-4264]{Mattias Lazda}
\affiliation{Dunlap Institute for Astronomy \& Astrophysics, University of Toronto, 50 St.~George Street, Toronto, ON M5S 3H4, Canada}
\affiliation{David A.~Dunlap Department of Astronomy \& Astrophysics, University of Toronto, 50 St.~George Street, Toronto, ON M5S 3H4, Canada}

\author[0000-0002-4209-7408]{Calvin Leung}
\affiliation{Miller Institute for Basic Research, University of California Berkeley, Berkeley, CA 94720, USA}
\affiliation{Department of Astronomy, University of California Berkeley, Berkeley, CA 94720, USA}

\author[0000-0002-4279-6946]{Kiyoshi W. Masui}
\affiliation{MIT Kavli Institute for Astrophysics and Space Research, Massachusetts Institute of Technology, 77 Massachusetts Avenue, Cambridge, MA 02139, USA}
\affiliation{Department of Physics, Massachusetts Institute of Technology, 77 Massachusetts Avenue, Cambridge, MA 02139, USA}

\author[0000-0002-5794-2360]{Dant\'{e} M. Hewitt}
\affiliation{Anton Pannekoek Institute for Astronomy, University of Amsterdam, Science Park 904, NL-1098 XH Amsterdam, the Netherlands}

\author[0000-0002-4560-5316]{Jacob Willis}
\noaffiliation

\author[0009-0003-1765-8845]{Alyssa Atkinson}
\affiliation{Dunlap Institute for Astronomy \& Astrophysics, University of Toronto, 50 St.~George Street, Toronto, ON M5S 3H4, Canada}
\affiliation{David A.~Dunlap Department of Astronomy \& Astrophysics, University of Toronto, 50 St.~George Street, Toronto, ON M5S 3H4, Canada}

\author[0000-0003-3772-2798]{Kevin Bandura}
\affiliation{Lane Department of Computer Science and Electrical Engineering, 1220 Evansdale Drive, PO Box 6109, Morgantown, WV 26506, USA}
\affiliation{Center for Gravitational Waves and Cosmology, West Virginia University, Chestnut Ridge Research Building, Morgantown, WV 26505, USA}

\author[0000-0002-2878-1502]{Shami Chatterjee}
\affiliation{Cornell Center for Astrophysics and Planetary Science, Cornell University, Ithaca, NY 14853, USA}

\author[0000-0001-6128-3735]{Nina V. Gusinskaia}
\affiliation{ASTRON, the Netherlands Institute for Radio Astronomy, Postbus 2, NL-7990 AA Dwingeloo, the Netherlands}
\affiliation{Anton Pannekoek Institute for Astronomy, University of Amsterdam, Science Park 904, NL-1098 XH Amsterdam, the Netherlands}

\author[0000-0003-2116-3573]{Adam E. Lanman}
\affiliation{MIT Kavli Institute for Astrophysics and Space Research, Massachusetts Institute of Technology, 77 Massachusetts Avenue, Cambridge, MA 02139, USA}
\affiliation{Department of Physics, Massachusetts Institute of Technology, 77 Massachusetts Avenue, Cambridge, MA 02139, USA}

\author[0000-0002-0772-9326]{Juan Mena-Parra}
\affiliation{Dunlap Institute for Astronomy \& Astrophysics, University of Toronto, 50 St.~George Street, Toronto, ON M5S 3H4, Canada}
\affiliation{David A.~Dunlap Department of Astronomy \& Astrophysics, University of Toronto, 50 St.~George Street, Toronto, ON M5S 3H4, Canada}

\author[0000-0003-0510-0740]{Kenzie Nimmo}
\altaffiliation{NASA Einstein Fellow}
\affiliation{Center for Interdisciplinary Exploration and Research in Astronomy, Northwestern University, 1800 Sherman Avenue, Evanston, IL 60201, USA}
\affiliation{Center for Astrophysics $|$ Harvard \& Smithsonian, 60 Garden St, Cambridge, MA 02138, USA}

\author[0000-0002-8912-0732]{Aaron~B.~Pearlman}
\altaffiliation{NASA Hubble Fellow.}
\affiliation{MIT Kavli Institute for Astrophysics and Space Research, Massachusetts Institute of Technology, 77 Massachusetts Avenue, Cambridge, MA 02139, USA}
\affiliation{Department of Physics, McGill University, 3600 rue University, Montr\'eal, QC H3A 2T8, Canada}
\affiliation{Trottier Space Institute, McGill University, 3550 rue University, Montr\'eal, QC H3A 2A7, Canada}

\author[0000-0002-4795-697X]{Ziggy Pleunis}
\affiliation{Anton Pannekoek Institute for Astronomy, University of Amsterdam, Science Park 904, NL-1098 XH Amsterdam, the Netherlands}
\affiliation{ASTRON, the Netherlands Institute for Radio Astronomy, Postbus 2, NL-7990 AA Dwingeloo, the Netherlands}

  \author[0000-0002-3430-7671]{Alexander W. Pollak}
  \affiliation{SETI Institute, 339 Bernardo Avenue, Suite 200 Mountain View, CA 94043, USA}

\author[0009-0005-6633-3945]{Gurman Sachdeva}
\affiliation{Dunlap Institute for Astronomy \& Astrophysics, University of Toronto, 50 St.~George Street, Toronto, ON M5S 3H4, Canada}
\affiliation{David A.~Dunlap Department of Astronomy \& Astrophysics, University of Toronto, 50 St.~George Street, Toronto, ON M5S 3H4, Canada}
  
\author[0000-0002-4823-1946]{Vishwangi Shah}
\affiliation{Department of Physics, McGill University, 3600 rue University, Montr\'eal, QC H3A 2T8, Canada}
\affiliation{Trottier Space Institute, McGill University, 3550 rue University, Montr\'eal, QC H3A 2A7, Canada}

\correspondingauthor{Shion Andrew}
\email{shiona@mit.edu}

\collaboration{99}{(CHIME/FRB Collaboration)}


\begin{abstract}
Precise localizations of fast radio bursts (FRBs) to specific environments within their host galaxies with very-long-baseline interferometry (VLBI) have provided valuable insights into the nature of their progenitors. To date, such localizations have relied primarily on traditional VLBI facilities, whose observational capabilities limit them to targeted follow-up of a small subset of repeating FRBs that are observable at frequencies above $\sim$1 GHz. CHIME/FRB Outriggers is a low-frequency (400--800 MHz), wide-field VLBI network designed to overcome these limitations, aiming to localize a substantial fraction of CHIME-detected FRBs to $\sim$50 mas precision. In this work, we detail the analysis procedure used to localize FRBs with the CHIME/FRB Outriggers VLBI array. With over a thousand test localizations consisting of continuum calibrators and pulsars with known VLBI positions, we provide a comprehensive demonstration of the VLBI array's ability to robustly meet, and under some conditions exceed, the nominal astrometric specification required for the central science goals of CHIME/FRB Outriggers. By demonstrating that a wide-field survey instrument can deliver routine $\sim$50 mas astrometry on one-off FRBs, this work lays the foundation for the upcoming generation of wide-field VLBI FRB surveys and the population-scale FRB samples they will enable.
\end{abstract}
\section{Introduction}
Fast radio bursts (FRBs) are millisecond-duration radio transients whose
luminous, coherent radio emission and extragalactic distances make them powerful
astrophysical and cosmological probes. Precise localizations have proven valuable to
nearly every aspect of FRB science: they enable host-galaxy associations and
redshift determinations \citep{Chatterjee_2017,Marcote_2017,DSA_110,ASKAP_2025,KKO_CAT_2025ApJS,Meerkat_2026}, situate FRBs in the context of their local environments \citep[e.g.][]{Tendulkar_2021ApJ,Nimmo_2022,Hewitt_2024,2024NatAs...8.1429C}, and turn FRBs into distance-calibrated probes of intervening
intergalactic and circumgalactic media via the dispersion, scattering, and
rotation measures imprinted on each burst \citep[e.g.][]{Macquart_2020,Ocker_2022ApJ,Glowacki_2025}. While arcsecond-scale localizations are sufficient to
identify host galaxies (and therefore obtain redshifts) for the majority of FRBs \citep{Eftekhari_2017},
$\sim$ milliarcsecond-scale localizations provided by very-long-baseline interferometry (VLBI) are required to resolve the kiloparsec and parsec scale environments around each source within its host. Such localizations offer a far more detailed and diverse picture of FRBs than what is accessible at arcsecond scales; for example, some FRBs have been associated with persistent compact radio emission and extreme magnetoionic environments \citep[e.g.][]{Marcote_2017, Moroianu_2026}, others have been localized hundreds of pc away from the nearest region of active star formation \citep{Tendulkar_2021ApJ}, and at least one has been pinpointed to a globular cluster in an otherwise typical star-forming galaxy \citep{Kirsten_2022}.

The increasing evidence pointing towards a non-monolithic set of FRB progenitors has necessitated a much larger sample of precisely localized FRBs to understand them on a population level. To date, samples of
$\mathcal{O}(100)$ FRBs have been associated with host galaxies through
$\sim$ arcsecond-scale localizations from connected-element interferometers such as
the Australian Square Kilometre Array
Pathfinder~\citep{ASKAP_2025} and the Deep Synoptic Array~\citep{DSA_110}, while only
$\mathcal{O}(10)$ well-studied repeating FRBs have been localized to
milliarcsecond precision, primarily through targeted follow-up with dedicated VLBI facilities \citep{Marcote_2022}. The Canadian Hydrogen Intensity Mapping Experiment/Fast Radio Burst (CHIME/FRB) Outriggers~\citep{outriggers_overview} VLBI network is the first of a
generation of upcoming FRB radio survey instruments \citep{vanderlinde2019canadian,lin2022burstt,casm} that connect wide-field instruments into a VLBI network for $\sim$ milliarcsecond resolution over a large sky area. The central science goal of CHIME/FRB Outriggers is to localize a substantial fraction of CHIME-detected FRBs (the majority of which are not observed to repeat) to within their host galaxies and
increase the $\sim 50$ milliarcsecond-localized sample by an order of magnitude, with this target set by the angular
resolution of the best multi-wavelength follow-up facilities (e.g.\ HST,
JWST, ALMA) whose resolutions range from $\sim$10--100\,mas. 

Achieving this angular precision with the CHIME/FRB Outriggers network for one-off FRBs, however, presents a distinct technical challenge relative to the VLBI localizations of repeating FRBs achieved to date. These localizations have been carried out with traditional VLBI facilities, primarily the European VLBI Network (EVN) \citep{Marcote_2017,Tendulkar_2021ApJ,Kirsten_2022,Nimmo_2022,Hewitt_2024,Moroianu_2026}, and benefit from several conditions that standard VLBI calibration procedures are built around. 
The positions of repeaters are known a priori to $\lesssim$ arcminute precision, allowing a phase-reference calibrator typically within $\sim$1$^\circ$ of the target to be selected in advance. VLBI repeater observations to date have also been conducted at $\gtrsim$ GHz frequencies, within the cm-wavelength regime in which the bulk of VLBI astrometry has been performed and where the residual ionospheric contribution to the phase is comparatively modest. Moreover, because the source repeats, observations are distributed over multiple epochs that are then combined to build up the final localization. Combined with the extensive $uv$-plane coverage provided by many baselines as a function of time, the large number of independent measurements allows temporal, source-specific, and antenna-specific systematics to be isolated and calibrated.

None of these conditions hold for one-off FRBs detected by CHIME/FRB Outriggers. CHIME detects FRBs at sky positions that are not known a priori within an instantaneous field of view of $\sim 200$\,deg$^2$, and because calibrators cannot be scheduled in advance, they must instead be drawn from whichever sources happen to lie in the primary beam at the time of detection. The calibrator density at the sensitivity of the Outriggers VLBI array ($\sim$100\,mJy for hundreds of ms integrations) limits the system to typical target--calibrator angular separations of tens of degrees \citep{2025_Andrew}, compared with $\sim$1$^\circ$ for typical EVN localizations. Furthermore, the 400--800~MHz observing band lies well below the frequencies at which standard VLBI is conducted; in conjunction with the large target--calibrator separations, this results in spatiotemporal ionospheric phase contributions that dominate the astrometric error on the $\gtrsim 10^3$~km baselines of the network. The Outriggers array also consists of only one sensitive core and three smaller Outrigger stations, and because multi-epoch observations of non-repeating sources are not possible, each localization rests on a small number of independent measurements. 

The diversity of CHIME/FRB detections compounds the calibration problem. In addition to calibration quality, the final astrometric precision depends on burst signal-to-noise (S/N) and effective bandwidth. The CHIME-detected FRB population spans orders of magnitude in fluence and a wide
range of fractional bandwidths \citep{CHIME_2026ApJS}: while bright, broadband bursts \citep[e.g.][]{RBFloat_2025} that provide high
cross-correlation S/N on every CHIME-Outrigger baseline are straightforward to calibrate, faint and/or narrowband sources are highly sensitive to the quality of the calibration solution. Because the Outrigger telescopes have a reduced collecting area relative to CHIME, the majority of CHIME-detected FRBs are faint in the VLBI cross-correlation.  Localizing a large fraction of CHIME-detected FRBs to high fidelity, therefore, requires a VLBI procedure that is robust, automated for the high CHIME/FRB event rate of $\sim$hundreds of VLBI localizations per year, and validated end-to-end against a control sample
of sources with independently known positions and similar brightnesses.

In this paper we present the astrometric procedure used to localize FRBs with the fully comissioned CHIME/FRB Outriggers VLBI array and provide a comprehensive demonstration of its performance. The remainder of the paper is organized as
follows. Section~\ref{sec:observational_context} provides an overview of the end-to-end localization and fringe-fitting procedure. Section~\ref{sec:calibration_dataset} describes the
direction-dependent calibration of our primary systematics. Section~\ref{sec:final} presents the final localization
procedure and the empirically calibrated error distribution from a sample
of more than a thousand test localizations of continuum calibrators and 
pulsars, 
demonstrating that the network is capable of meeting its 50~mas astrometric specification
for a substantial fraction of CHIME-detected FRBs. 

\section{Procedure and Visibility Model}
\label{sec:observational_context}
CHIME and its Outriggers are each stand-alone interferometers 
operating from 400--800\,MHz with a shared primary-beam footprint of
$\sim\!2^\circ\,\mathrm{(EW)}\times\!100^\circ\,\mathrm{(NS)}$. CHIME
contains 1024 dual-polarized feeds; its \kkoname\ Outrigger (KKO) contains
64; and its Outriggers in Green Bank (GBO) and Hat Creek (HCO) each contain
128 \citep{lanman2024kko,outriggers_overview,hco_paper}. CHIME and KKO form the shortest of the three CHIME-Outrigger baselines (66km), and the two
roughly orthogonal and longest baselines --- CHIME-GBO (3334km predominantly
east--west) and CHIME-HCO (955km predominantly north--south) --- 
dominate the astrometric precision and constrain the source position in two dimensions \citep[see Figure 2 of][]{outriggers_overview}. While the Outrigger-Outrigger baselines in principle could also be used,
in practice our observed calibrators (as well as the vast majority of FRBs) are not bright enough to exceed the noise threshold on the Outrigger-Outrigger baselines due to the limitations of the current system, which cap the integration times on our VLBI observations to $\sim$100~ms \citep{2025_Andrew}. We note that unlike traditional VLBI
instruments designed for imaging, CHIME/FRB Outriggers array functions primarily as
an astrometric VLBI array that localizes unresolved point sources (i.e. FRBs) without the need to reconstruct spatial brightness distributions. Although the network has minimal instantaneous $uv$ coverage per snapshot observation, the requirement that only a single 2D Fourier mode be measured, in conjunction with CHIME's large fractional bandwidth which suppresses the 
sidelobe response of the synthesized beam\footnote{In ``canonical'' imaging terms, 
$uv$ points are added radially with wider bandwidth \citep{thompson2017interferometry1}.}, turns the network's baseline configuration into a theoretically manageable constraint.

\subsection{Procedure overview}
Here, we provide a high-level overview of the full end-to-end localization procedure used by CHIME/FRB Outriggers for full-array baseband (channelized voltage) observations, now systematized and automated for the complete VLBI array; additional details on the fringe-fitting and calibration procedures are provided in Section \ref{sec:visibility_model}.

When the CHIME/FRB search backend detects a radio transient, triggers are sent to CHIME and all its Outriggers to record a $\sim$100~ms ``snapshot'' of baseband data (1024 spectral channels each with a resolution of 390\,kHz) capturing the burst. The basic workflow following this full-array baseband data acquisition and transfer to a central processing site 
consists of the steps detailed below.

\subsubsection{Station beams}
Since each full-array baseband snapshot covers the telescope's full $\sim 200$ square degree primary beam, we use multiple phase center beamforming to form individual beams at the CHIME core and each Outrigger station toward a set of pointings which consist of astrometric calibrators in the field of view in addition to the target. The initial target pointing is provided by a CHIME-only diffraction-limited ($\sim$arcminute) localization provided by an offline interferometric localization pipeline \citep{Michilli_2021}. Because the Outrigger station beams are wider than CHIME's by a factor of $\sim 4$ in the east--west direction and $\sim 2-4$ in the north--south direction \citep{lanman2024kko,outriggers_overview,hco_paper,gbo_paper} an $\sim$ arcminute pointing error is a small enough fraction of the Outrigger beams that the signal-to-noise loss incurred at the beamforming stage is negligible ($\lesssim 1\%$; see Equation~37 of \citealt{Masui_2017}).

\begin{figure}
    \centering
    \includegraphics[width= \linewidth]{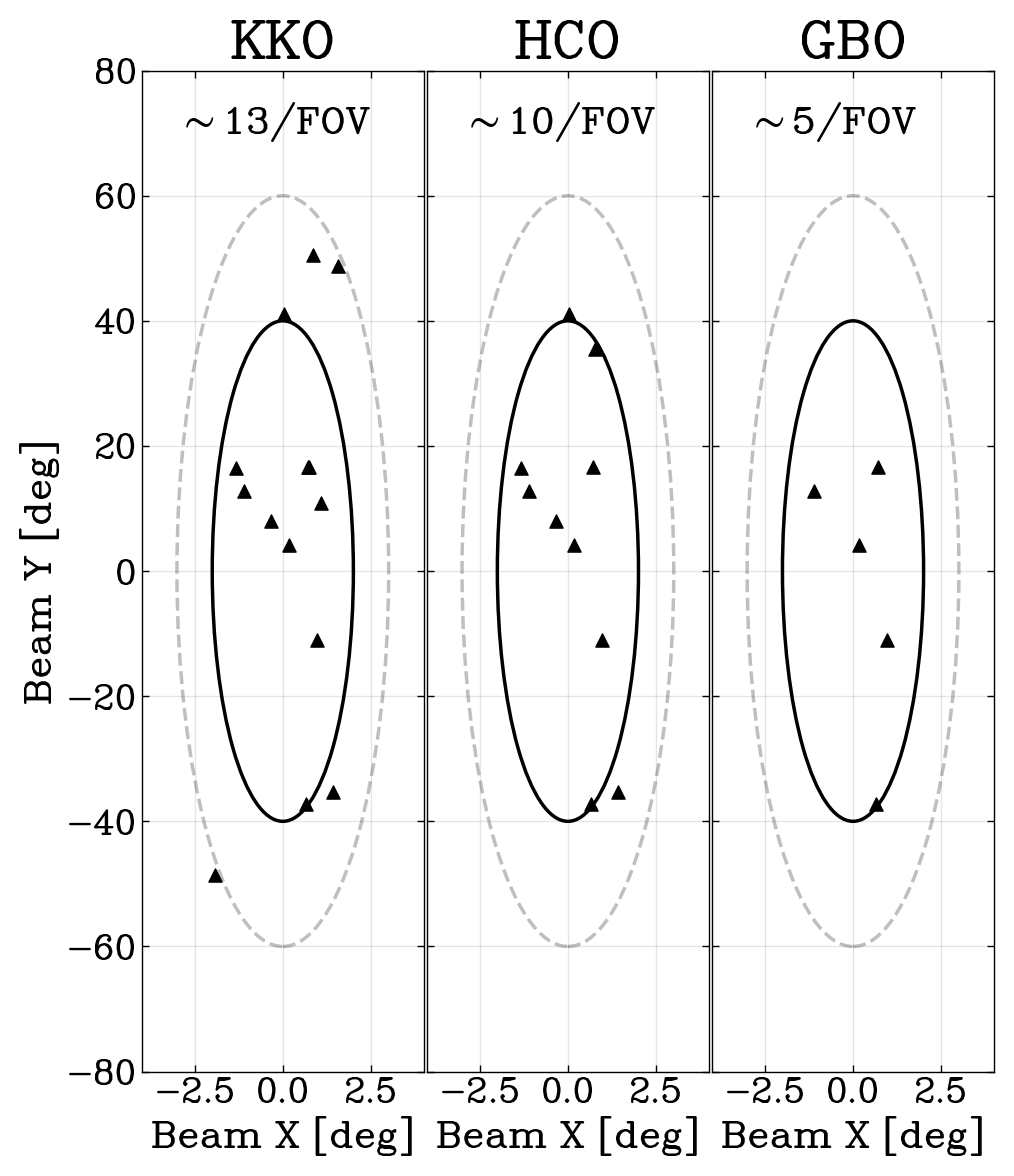}
    \caption{Typical calibrator density on the CHIME-KKO ($\sim 60$km), CHIME-HCO ($\sim 1000$km), and CHIME-GBO ($\sim 3000$km) baselines. 
    The calibrators are plotted in topocentric beam coordinates relative to CHIME zenith, and the dashed gray line encompasses $\sim$95\% of all 
    CHIME-FRB detections. }
    \label{fig:calibrator_FOV}
\end{figure}

Calibrators are selected from the Radio Fundamental Catalog (RFC) ~\citep{petrov2021wide,Petrov_2025} for which milliarcsecond astrometry is available. Snapshot durations correspond to a sensitivity of $\sim$ 100mJy in CHIME-Outrigger visibilities and a calibrator density of $\sim 4-10$ calibrators on our longest baselines \citep{2025_Andrew}: the typical calibrator density
across our field of view on all CHIME-Outrigger baselines is shown in 
Figure \ref{fig:calibrator_FOV}. Radio frequency interference (RFI) is also spatially filtered at the beamforming stage using the algorithm described in \cite{Andrew_2026}. At the conclusion of this stage, ``multi-beam'' files consisting of all frequencies and pointings are saved to disk.
\subsubsection{Calibrator fringe finding}
For calibrator pointings, the voltage data at each station is delay-compensated and integrated over the full duration of the baseband capture using ~\corrname~\citep{leung2024vlbi}. In addition to spatially filtering interference, we calculate the spectral kurtosis~\citep{nita2020generalized} from the full-array baseband data in each snapshot as a function of frequency at the native resolution of 390~kHz, and additional outliers above $m$ median absolute deviations are flagged to reject RFI, where $3 < m < 10$ is optimized for each pointing independently to maximize the S/N of the calibrator fringes. 

Across the near-octave bandwidth on the longest CHIME--Outrigger baselines, the differential ionospheric slant TEC (dSTEC) can cause substantial phase decoherence if not removed. Following \citet{2025_Andrew}, after forming cross-correlated visibilities we search a grid of dSTEC values and, at each trial, Fourier transform the visibilities over frequency to search a grid of delays to confirm the presence of fringes for each calibrator on a given baseline.

\subsubsection{Target correlation and pulse profiling}
\label{sec:gating}
For the single target pointing, the voltage data from each station is delay-compensated, coherently dedispersed,  integrated around the burst profile, and cross-correlated over time between stations to form visibilities. The post-fringestopping and coherent-dedispersion steps can be described by the operation:
\begin{equation}
    V_{AB}(\nu,\Delta \ell) = \sum_t w_{AB}(t,\nu)\,b_A(t,\nu)\,b_B^*(t+\Delta \ell,\nu),
\end{equation}
where $V_{AB}$ are the visibilities formed between stations $A$ and $B$, $b_A(t,\nu)/b_B(t,\nu)$ is the de-dispersed beamformed baseband data at station $A/B$ (suppressing the pointing dependence for clarity), $w_{AB}(t,\nu)$ a complex weight applied to the correlation product, $t$ the time sample, $\nu$ the frequency, and the integer lag $\Delta\ell$ scans the residual delays examined during the subsequent fringe search. Denoting the signal component of the baseband data at station $A$ as $s_A(t,\nu)$, in white noise\footnote{In the case of time-variable interference, the more general form of the optimal weights requires an estimate of the noise covariance matrix \citep{Andrew_2026}. While in principle this could be obtained from the unbeamformed baseband data, we defer the detailed implementation and testing of such a method to a future work.} the integrated signal-to-noise is maximized by the matched-filter weight
\begin{equation}
    w_{AB,\rm{opt}}(t,\nu) \propto \left(s_A(t,\nu)\,s_B^*(t,\nu)\right)^* = |s(t,\nu)|^2\,e^{-i\phi_{AB}(t,\nu)},
\end{equation}
which factors into the burst intensity $|s(t,\nu)|^2$ and a relative phase $\phi_{AB}(t,\nu)$ between the two stations. The relative phase $\phi_{AB}(t,\nu)$ contains only a residual time-dependent geometric delay between $A$ and $B$; after fringestopping to the initial pointing, the correlator has already removed the bulk of this delay \citep{leung2024vlbi} and $\phi_{AB}(t,\nu)$ is approximately constant over the integration. $w_{AB}(t,\nu) \propto |s(t,\nu)|^2$ is therefore primarily a time-dependent amplitude correction---in the simplest case a constant (unit) value across the on-pulse window.

\begin{figure}
    \centering
    \includegraphics[width= \linewidth]{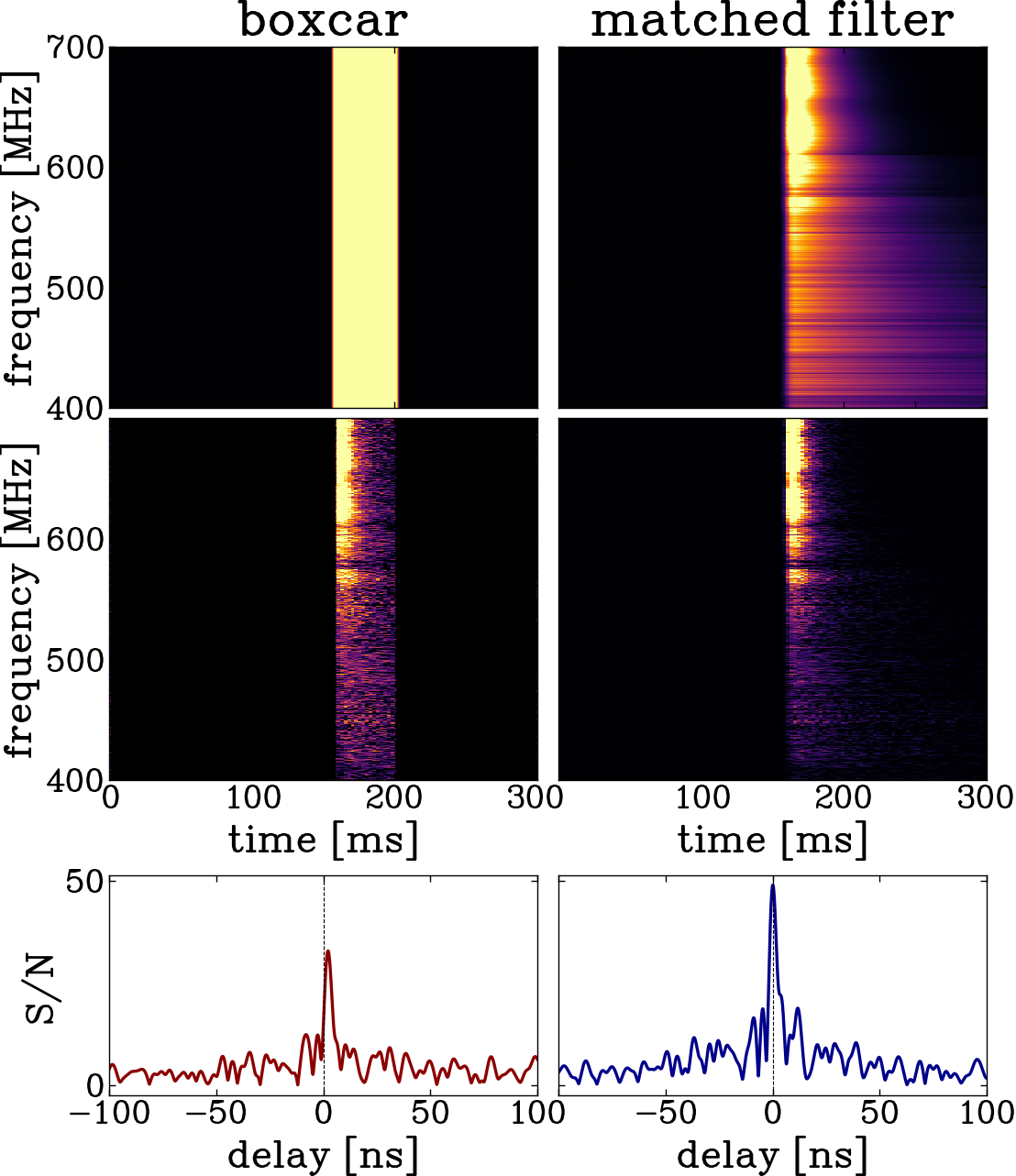}
    \caption{Performance comparison using a naive boxcar gate (left)
    and a matched filter (right). The boxcar width was chosen to maximize the cross-correlation signal given the constraint on the shape of the filter (over many width and start time trials). The top panel compares boxcar to matched filter weights, and the middle panel shows corresponding waterfall plots on CHIME data after applying the weights. The bottom panel shows cross correlated signal to noise with respective 
    weights applied.}
    \label{fig:fitburst_gating}
\end{figure}

However, an additional signal-to-noise boost can be provided when the pulse profile is known or can be modeled \citep{brisken2010difx}. This is particularly useful for the CHIME Outriggers network because of its strongly asymmetric distribution in collecting area across telescopes: the signal-to-noise is largest at CHIME and the pulse can be best fit directly from the CHIME autocorrelation data using morphological models \citep[e.g.][]{Fonseca_2024}.  An example of the improvement in performance when using signal weights instead of a simple box-car weight on CHIME-Outriggers data for an FRB is shown in Figure \ref{fig:fitburst_gating}, 
where weights $w_{AB}(t,\nu)$ are fit to the CHIME autocorrelated baseband data as a function of frequency and time.

In general, the improvement over a boxcar gate is larger the more poorly the boxcar approximates the true signal, as is the case for FRBs with multiple components, complex frequency structure, or large scattering tails. The full time--frequency matched filter is particularly important for recovering signal at low frequencies (where scattering is also the strongest), since this low-frequency signal is in turn needed to break the degeneracy between the ionospheric and geometric delay phases (see Section \ref{sec:phase_calibration}). 

The discussion so far has emphasized the signal-to-noise boost from the time variability of the weights, but the frequency-dependent amplitude of the visibilities are also required for the fringe fit. For a narrowband burst, for example, a matched filter along the frequency axis outperforms subbanding the data (a ``boxcar'' in along the frequency axis), and helps recover signal when the burst is too faint to estimate the visibility amplitudes from the cross-correlated data directly (detailed further in Section~\ref{sec:amp_fitting}).

\subsubsection{Target calibration and fringe fitting}
Given the $\sim$arcminute precision of the initial correlator pointing provided by the CHIME-only interferometric localization \citep{Michilli_2021}, the target fringe must be located over a corresponding range of residual delays. In principle this search can be distributed over a grid of $M$ initial phase centers, each covering a fraction $1/M$ of the delay range near zero lag; this is the approach preferred for long-integration targets, where re-reading and delay-compensating the full dataset dominates the cost, so that forming additional phase centers within a single pass is comparatively cheap. For a gated single pulse, however, only the on-pulse ($\sim 10$\,ms) samples enter the correlation, so recorrelation is inexpensive: we instead correlate to a single phase center and upchannelize on the longest baselines to search for fringes on ``off-lags'' ($|\tau_{\rm{geo}}|>2.56\,\mu$s $\leftrightarrow \sim 0.5-1'$; see also~\citealp{leung2024vlbi}).

Since the systematic contributions to the measured visibility phases (detailed further in Section \ref{sec:visibility_model}) are baseline dependent, both calibration and fringe fitting is carried out independently for each CHIME-Outrigger baseline. Calibrator fringes are identified for each baseline following the procedure in \cite{2025_Andrew}, and target visibilities are directly phase referenced to all calibrators on a given baseline ($M$ separate calibrated visibilities) to remove all time dependent contributions to the phase. 

We adopt the point-source visibility model detailed in Section \ref{sec:visibility_model} to perform the fringe fit on our calibrated visibilities for each baseline. The initial localization for a burst is obtained by maximizing the posterior probability of our visibility model (Eq.~\ref{eq:loglike_marg}) over the residual delay $\tau$, the differential slant TEC, and the polarization-dependent amplitude scaling $s_\alpha$ (Section~\ref{sec:amp_fitting}), with the ionospheric prior taken from the thin-shell fit to the in-beam calibrators (Section~\ref{sec:thin_shell}). The $N$ calibrated visibilities on each baseline are combined through the joint-calibrator likelihood (Section~\ref{sec:joint_cal_lik}), and a source is considered ``detected'' if the final detection significance (defined in Appendix~\ref{appendix:snr}) exceeds $5\sigma$. The delay posteriors on each baseline are convolved with an error kernel (see Section \ref{sec:superresolve}) to add remaining systematic contributions ``in quadrature" to the localization uncertainty.

\subsubsection{Iterative recorrelation and final localization}
For each baseline, the delay posteriors are mapped to localization contours with \texttt{pycalc11} \citep{pycalc}, which implements the consensus delay model \texttt{CALC} used in VLBI. The localization contours on each baseline are combined independently, and the best fit position is taken as the centroid of the resulting 2D sky contour. Because the residual delay rate $\dot\tau$ acting over the dispersive sweep due to an incorrect initial correlation position produces a $\propto 1/\nu$ phase degenerate with an ionospheric phase \footnote{This can be readily seen from evaluating the second term of the first order Taylor expansion $\tau_{geo}(\nu)=\tau_{geo}(\nu_0) +\dot{\tau}(t(\nu_0))\bigg[t(\nu_0)-t(\nu)\bigg]$ where $t(\nu_0)-t(\nu)\propto \frac{\rm{DM}}{\nu^2}$ contains the DM sweep}, the target visibilities are recorrelated to the best initial localization position to improve the accuracy of the ionospheric prior by suppressing the residual delay rate across the band. All the aforementioned steps are repeated iteratively with the refined target position until convergence ($\Delta \tau <0.1\sigma_{\rm{\tau,loc}}$) to provide the final localization contour.

\subsection{Point Source Visibility Model} 
\label{sec:visibility_model}
Now, we detail our point source visibility model, which describes the systematics present in our data down to the $\sim$ 10~mas level of precision. Additional systematics which matter at the sub-radian level across the 400-800 MHz band are discussed in Section~\ref{sec:superresolve}; however, because these are not relevant for the \emph{primary} science goal of the CHIME/FRB Outriggers survey, we defer a detailed treatment of these systematics to a future work. 

For a point source with true position $\hat{n}$ initially fringestopped to $\hat{n}'$ phase referenced to an unresolved in-beam calibrator at position $\hat{c}$, the most general form of the signal-component of the visibilities formed between stations $A$ and $B$ at frequency $\nu$ in polarization hand $\alpha$ (of the four possible pairs of polarization combinations, we use the two co-polarized linear polarization combinations between stations) in the narrowband limit can be decomposed into an amplitude and a phase
\begin{equation}
    \label{eq:basic_v_model}
    V_\alpha^{AB}(\nu,\hat{n},\hat{c})=S_\alpha^{AB}(\nu,\hat{n})e^{i \phi_{AB,\rm{sys}}(\nu,\hat{n},\hat{c})} e^{2\pi i \nu \tau_{\rm{geo}}(\delta \hat{n},\vec{b}_{AB})}
\end{equation}
where $\delta \hat{n}\equiv \hat{n}^{'}-\hat{n}$, $\vec{b}_{AB}$ denotes the baseline vector between stations $A$ and $B$, $\tau_{geo}$ the residual geometric delay containing the true position of the target, and $\phi_{AB,\rm{sys}}$ the unwanted systematic contribution to the phase. Here, we have assumed $\delta \hat{n}$ to be small enough that the residual pointing error manifests as a temporally constant overall phase: note, however, that we have omitted the second order contribution due to the fringe rate, which arises from the time evolution of $\vec{b}_{AB}$ over the frequency due to the dispersion sweep. The magnitude of this term is at the $\sim$ radian level on the longest baselines of CHIME/FRB Outriggers for typical FRB DMs and pointing errors \citep{leung2024vlbi}, and because it is completely degenerate with the ionospheric contribution to the phase (see Section \ref{sec:phase_calibration}) we choose to omit it from the fringe fit. Although this introduces a small ($\sim$ radian) initial error in the ionospheric prior, it is easily corrected for upon recorrelation to the initial fringe fit position.

\subsubsection{Phase model}

\label{sec:phase_calibration}
Because our calibrators are observed contemporaneously with our FRBs and our phases are observed to be stable over the full duration of the baseband capture ($<1\,\mathrm{s}$) at $\ll$ 1 radian for even our longest baseline \citep{2025_Andrew}, phase referencing removes all contributions to the systematic phase that are both time dependent and spatially invariant (e.g. differential clock delays). We consider two main contributions to the remaining systematic contribution to the phase, $\phi_{AB,\rm{sys}}$, both of which primarily depend on the relative location of the target to the calibrator.

The first we term a ``baseline offset'': an error in the assumed three-dimensional position of an Outrigger station. In conventional ``fully commissioned'' astrometric VLBI networks this contribution is negligible, since station coordinates have already been tied to the International Terrestrial Reference Frame at the millimeter level through decades of dedicated geodetic VLBI \citep{Geod_Petrov}. At the start of the commissioning process, the Outrigger station positions were only known through local GPS measurements, accurate to tens of meters \footnote{the GPS datum taken at each site is offset from the array center by roughly 10s of meters, while the GPS position itself is accurate to about a meter}, which translates to an astrometric error of several arcsec on our longest baseline (3000 km). We therefore measure the station positions with a calibrator dataset ($\delta \hat{n}$ in Equation \ref{eq:basic_v_model} is zero) by modeling this contribution to the phase as an overall delay:
\begin{equation}\label{eq:baseline_offset}
    \Delta\delta\tau_B = \delta\vec{B}\cdot(\hat{n}-\hat{c})/c,
\end{equation}
where $\delta\vec{B}$ is the station-position error. Because it depends only on the fixed station geometry, this global calibration is time-independent---in contrast to the ionospheric calibration below---and need only be performed once (analysis detailed in Section~\ref{sec:baseline_offset}). In the fringe model applied to FRBs we therefore omit this term, assuming that the residual baseline-offset delay after calibration is $<1$\,ns, which we validate in Section~\ref{sec:baseline_offset}.

The other major systematic contribution to the measured visibility phase is differential delays introduced by the ionosphere, which are typically $\sim$ tens of nanoseconds (hundreds of mas) on our longest baselines. While the bulk of the differential ionosphere between CHIME and each outrigger station is removed via contemporaneous phase referencing, there is a remaining direction-dependent residual due to the fact that the calibrator is often more than several degrees away from the target. This provides the final phase model post referencing  \citep{outriggers_overview,leung2024vlbi}:

\begin{equation}\label{eq:fringe_model}
    P_\nu(\tau, \rmtec) =
    \exp\!\left[2\pi i \left(\nu\tau + \frac{\kdm \cdot \Delta s\rmtec}{\nu}\right)\right]
\end{equation}
with $\kdm = 1344.54\;\mathrm{MHz\;TECU^{-1}}$, $\tau$ the geometric delay, and 
$\Delta$sTEC the differential slant TEC between the calibrator and target across stations. We treat $\Delta$sTEC as a nuisance parameter and marginalize over it numerically with a calibrator-derived prior (see Section \ref{sec:thin_shell}). Unlike GHz VLBI astrometry, where the residual ionosphere is a small enough systematic contribution to be folded into the error budget in quadrature, the complex interplay between  bandwidth, signal-to-noise and ionosphere prior in affecting the uncertainty in the final source position makes it difficult to calibrate error bars ``ad-hoc''. A central advantage of a visibility model of this form is that because the differential ionosphere is fit jointly with the geometric delay, its covariance with the source position propagates directly into the localization uncertainty, and these dependencies are captured self-consistently in the posterior.

\subsubsection{Amplitude model}
\label{sec:amp_fitting}

Following our visibility model in Equation \ref{eq:basic_v_model}, the optimal estimator assuming Gaussian noise is given by the log-likelihood
\begin{equation}
    \label{eq:general_loglik}
    \ln \mathcal{L} = -\sum_{\nu,\alpha} \frac{\left|V_{\nu \alpha} - S_{\nu \alpha} P_\nu(\hat{n})\right|^2}{2 \sigma_\nu^2}
\end{equation}
with the phase term given by Equation \ref{eq:fringe_model}. The correlated visibility output of the correlator is computed as a function of frame lag (residual delay in units of 2.56$\mu$s between the post-fringestopped baseband data at CHIME and the Outrigger) with the correlated signal confined to a few lags around the fringe peak. We compute the errors on the visibilities $\sigma_\nu$ for both the calibrator and target empirically by taking the rms of the signal-free ``off-lags'' (or equivalently, the sky background surrounding the target).

While calibration on the phases––which encode the source position––are typically emphasized in astrometry, the accuracy of the relative amplitude weighting $S^\alpha_\nu$ also matters to the integrity of the fit. This is particularly true for FRB bursts, which typically exhibit significant spectral structure over the CHIME observing band. In principle, this spectral structure is encoded in the
measured visibility amplitudes $|V_\nu|$. However, for the signal to dominate
the visibility amplitude, the per-channel S/N must be high, which is
not the case for the vast majority of FRBs we observe; in most
cases the per-channel visibilities are noise-dominated and $|V_\nu|$ is a poor estimator
of the true amplitude. 

Instead, we estimate $S^\alpha_\nu$ from the CHIME 
autocorrelated data (total-power spectrum), which due to CHIME's larger
collecting area relative to its outrigger telescopes has a 
signal-to-noise larger by a factor of $\sim$2 relative to 
the cross-correlated visibilities. We model the amplitude as $S^\alpha_\nu = S_\nu\,s^\alpha$, separating the total-intensity spectral shape $S_\nu$ from the source's polarization state $s^\alpha$ \footnote{Since $S_\nu$ is measured from the polarization-summed autocorrelation, it is independent of the (Faraday-rotated) polarization state, so the factorization holds to leading order, with any residual frequency dependence of $s^\alpha$ absorbed into the polarimetric model}. We take the model for the burst spectrum that is fit to the autocorrelation in the gating procedure (see Section \ref{sec:gating}) as the spectral template $S_\nu$, and can either measure it via a polarimetric analysis of the autocorrelation data or directly fit out the polarization-dependent scaling $s^\alpha$ (which in contrast to $S_\nu$ is feasible since $N_\nu>>N_\alpha$).

Note that the \emph{measured} visibility amplitude is not simply the burst spectrum: it also carries the product of the two stations' beam responses toward the source, i.e.\ a differential bandpass between CHIME and each Outrigger \citep{beam_holography}. On large spatial scales the shape of the beam is a relatively smooth envelope, and because the Outriggers are oriented to observe nearly the same drift-scan strip \citep{outriggers_overview}, this overall beam shape is closely matched between CHIME and each Outrigger \citep{lanman2024kko}. Within that envelope, however, the beam is strongly chromatic and spatially varying: holographic measurements of CHIME's beam \citep{beam_holography} show channel-to-channel fluctuations of up to a factor of $\sim 2$ that depend on source position. This fine structure does not match between stations---residual differences arise from the $\sim 10^\circ$ zenith-angle offset between CHIME and the Outrigger sites, producing sensitivity mismatches that are not yet well modeled at the Outrigger stations.

Rather than trying to correct for this differential bandpass explicitly, we absorb it into the amplitude uncertainty. Assuming Gaussian priors centered on the measured
values, and writing $R_{\alpha\nu} \equiv \mathrm{Re}[V_{\alpha\nu}\,\bar{P}_\nu]$ for the real
part of the phase-derotated visibility, 
marginalizing over the amplitude nuisance parameters provides (see Appendix~\ref{appendix:joint_cal_lik_marg} for full derivation)

\begin{align}
    \label{eq:loglike_marg}
\ln\mathcal{L}_\mathrm{marg} &\propto
-\frac{1}{2}\sum_{\alpha,\nu}
\Bigg[
    \frac{\bigl(R_{\alpha\nu} - s_\alpha\,\bar{S}_\nu\bigr)^2}
         {\sigma_\nu^2 + s_\alpha^2\,\delta S_\nu^2}
    - \frac{R_{\alpha\nu}^2}{\sigma_\nu^2} \nonumber \\
&\qquad
    + \ln\!\left(1 + \frac{s_\alpha^2\,\delta S_\nu^2}{\sigma_\nu^2}\right)
\Bigg] \nonumber \\
&\qquad
    + \sum_{\alpha,\nu}
    \ln\!\left[\tfrac{1}{2}\,\mathrm{erfc}\!\left(-\frac{z_{\alpha\nu}}{\sqrt{2}}\right)\right]
\end{align}
with
\begin{equation}
z_{\alpha\nu} =
\frac{s_\alpha\,R_{\alpha\nu}\,\delta S_\nu^2 + \bar{S}_\nu\,\sigma_\nu^2}
     {\sigma_\nu\,\delta S_\nu\,\sqrt{\sigma_\nu^2 + s_\alpha^2\,\delta S_\nu^2}}
\end{equation}
where $\delta S_\nu$ is the amplitude uncertainty. In our case, $\delta S_\nu$ is typically dominated by the differential bandpass between CHIME and each outrigger station (which we take to be a factor of 2 due to the structure of the CHIME beam).

\begin{figure}[h]
    \includegraphics[width= \linewidth]{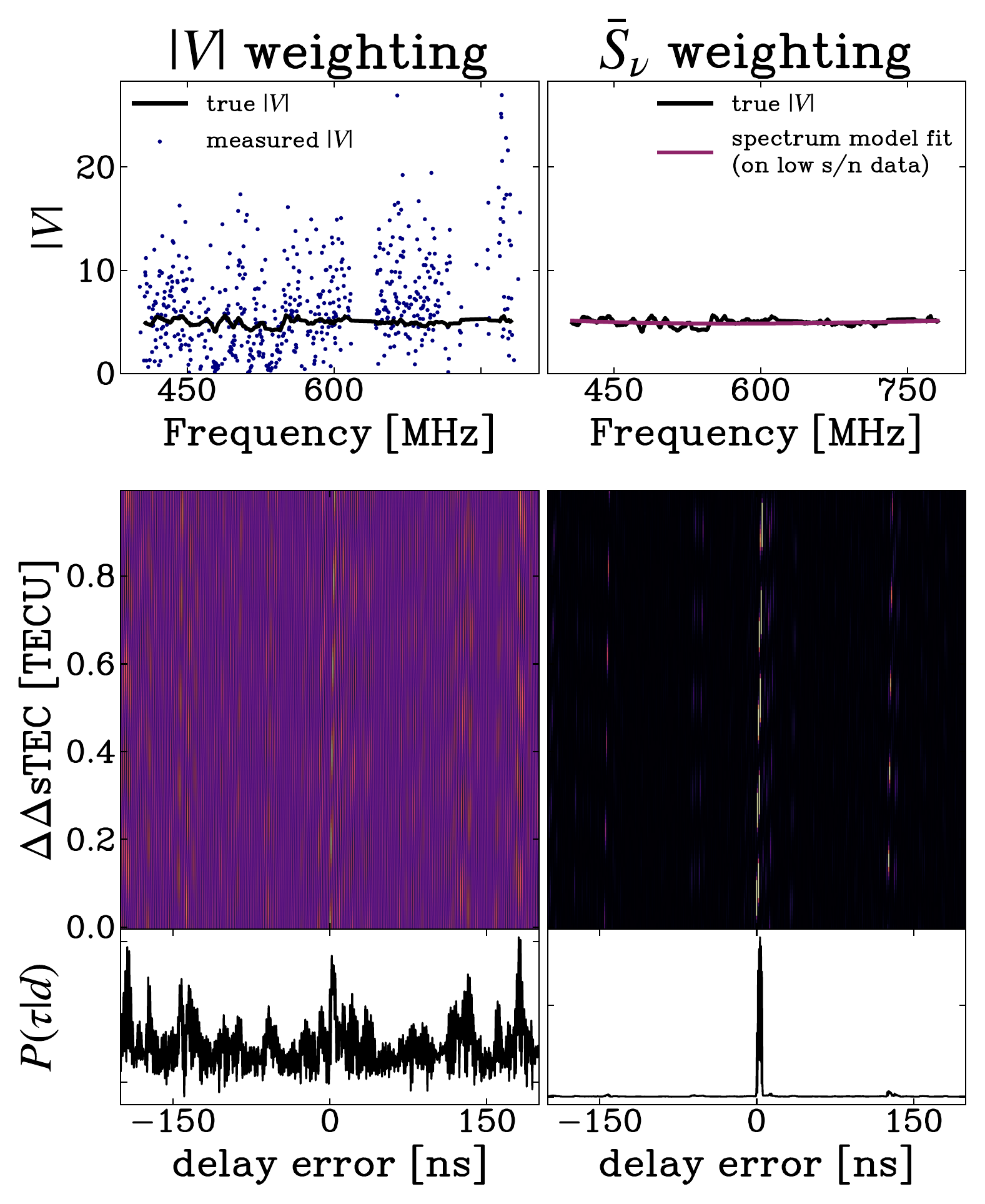}
    \caption{Visual demonstration of the improvement in localization error seen by adopting amplitude weighting described by Eq.~\ref{eq:loglike_marg} ($S_\nu = \bar{S}_\nu$) as compared to the naive ``bright point source" amplitude model given by Eq. \ref{eq:general_loglik} with $S_\nu = |V_\nu|$. We use cross correlated visibilities for the pulsar B0329+54 for this demonstration, since its VLBI position is known \citep{Kumar_2025} and when beamformed normally, this source is bright enough in the cross correlated visibilities to dominate the measured amplitude, which allows us to visually compare our adopted model values $S_\nu$ after degrading the S/N to the ``true" underlying spectrum. We mimic a low signal to noise measurement by decreasing the number of inputs used in the beamforming step (decreasing the effective collecting area of the individual stations). Top panels: measured spectrum of the ``noisy" spectrum (left), compared to the modeled spectrum fit to the CHIME-autocorrelated data (using the ``noisy gate''). The ``true" spectrum is overplotted in both figures, where we see a much better match with the model fit. The rippling feature in the ``true" spectrum arises from the differential beam response between CHIME and its Outriggers, which is the dominant contribution to the width of the prior uncertainty $\delta S_\nu$.  Middle panels: Marginalized fringe-fit posteriors (Equation \ref{eq:loglike_marg}) as a function of $\Delta \Delta $sTEC and delay. Bottom panels show the fringe fit posteriors in the middle panel after marginalizing over ddTEC. We find that folding
    in the model-derived amplitude template recovers sensitivity that
    would otherwise be lost to noise-dominated channels. }
    \label{fig:plot6_amp_weighting}
\end{figure}

An example demonstrating the performance of this likelihood with real data relative to amplitude weightings dictated directly by the measured cross correlated visibilities ($S_\nu = |V_\nu|$) is demonstrated in Figure \ref{fig:plot6_amp_weighting}. Without an external amplitude model the likelihood is extremely noisy and shows no discernible signal, whereas modeling $S_\nu$ recovers a clear detection. This is because $|V_\nu|$ is both a noisy and a biased amplitude estimate: for the low-S/N bursts typical of our sample, most channels are noise-dominated, and weighting by $|V_\nu|$ therefore assigns substantial weight to channels that contain only noise. Supplying an independent template $S_\nu$ from the high-S/N CHIME autocorrelation instead acts as a matched filter across frequency, concentrating the fit on the channels that actually contain burst flux and down-weighting the empty ones. This is especially important for spectrally structured or narrowband bursts, where the flux is confined to a fraction of the band and a self-derived weighting buries it beneath the noise-dominated channels.

\subsection{Joint-calibrator likelihood}
\label{sec:joint_cal_lik}
Our fixed baseband capture length means that in a given channel the calibrator signal-to-noise does not always exceed that of the target. In the limit that the calibrator noise dominates, the same target can be calibrated against $N_{\mathrm{cal}}$ in-beam calibrators to produce $N_{\mathrm{cal}}$ measurements that can be combined independently to improve the localization uncertainty. In the limit that the target noise dominates, however, the statistical uncertainties on the calibrated visibilities cannot be treated independently and a single calibrator (typically the brightest or the closest) must be chosen instead.

CHIME/FRB Outriggers data often occupies the intermediate regime,  where there is partial but not complete correlation between the resulting calibrated visibilities. We therefore 
explicitly model the covariance matrix between all pairs of  pointings. We adopt a phase-only calibration, $V_{\mathrm{cal},c} = V_{\mathrm{tar}}\,\overline{g_c}$ with $g_c = V_{\mathrm{calibrator},c}/|V_{\mathrm{calibrator},c}|$, so each calibrated visibility is the same target visibility with a bulk phase correction. To first order, the fluctuation of $V_{\mathrm{cal},c}$ then decomposes into a target-noise contribution rotated by $\overline{g_c}$---shared across calibrators---and an independent per-calibrator phase noise of amplitude $|V_{\mathrm{tar}}|$. Stacking the $N_{\mathrm{cal}}$ calibrated visibilities at a given (frequency $\nu$, polarization) into a vector, the covariance is therefore
\begin{equation}\label{eq:cov_matrix}
    C \;=\; D \;+\; \sigma_{\mathrm{tar}}^{2}\,\mathbf{w}\mathbf{w}^{\dagger},
\end{equation}
with $\mathbf{w} = (\overline{g_1},\dots,\overline{g_{N_{\mathrm{cal}}}})^{\top}$ the vector of unit calibrator phasors that rotate the shared target-noise fluctuation into each calibrated visibility, and
\begin{align}
    D_{cc'} &= \delta_{cc'}\,\frac{|V_{\mathrm{tar}}|^{2}\,\sigma_{\mathrm{cal},c}^{2}}{|V_{\mathrm{calibrator},c}|^{2}}
\end{align}
the diagonal matrix of independent per-calibrator noise.
The mean model for calibrated visibility $c$ at channel $\nu$ and polarization $\alpha$ is
\begin{equation}\label{eq:joint_model}
    m_{c,\nu\alpha} = s_\alpha\,S_\nu\,\exp\!\left[2\pi i\!\left(\nu\tau + \frac{\kdm\,\mathrm{TEC}_c}{\nu}\right)\right],
\end{equation}
with a shared delay $\tau$ and amplitude calibration $s_\alpha$ across calibrators, but a per-calibrator ionosphere $\mathrm{TEC}_c$ that absorbs the differential sTEC between the target and each calibrator's line of sight. Stacking the measured visibilities into a vector $\bf{y}$, we have the combined log-likelihood
\begin{equation}\label{eq:joint_loglik}
    \ln \mathcal{L} = -\tfrac{1}{2}\sum_{\nu,\alpha}(\mathbf{y_{\nu\alpha}}-\mathbf{m_{\nu\alpha}})^{\dagger}\,C^{-1}\,(\mathbf{y_{\nu\alpha}}-\mathbf{m_{\nu\alpha}})
\end{equation}
is evaluated per channel with frequencies and polarizations are taken as independent measurements. We derive the form of Equation \ref{eq:joint_loglik} after marginalizing over $S_\nu, s_\alpha$ in Appendix \ref{appendix:joint_cal_lik_marg}.

\subsection{Subradian systematics}
\label{sec:superresolve}
We note that the visibility model introduced in Section~\ref{sec:visibility_model} begins to break down beyond $\sim 10$\,mas for several reasons. The residual neutral atmosphere contributes a non-dispersive differential delay that phase referencing does not fully remove, particularly for larger target--calibrator separations; since their absolute positions (at $\sim $ mas level) are only available at GHz frequencies, the calibrator core shift (the frequency-dependent offset of the synchrotron self-absorbed core along the jet) \citep{Plavin_2019} biases the assumed calibrator position at the mas level; the beam-induced phase differs between CHIME and the Outrigger and varies across the field of view; and the ionosphere is plausibly non-stationary over the dispersive sweep, so a single differential slant TEC term no longer fully captures its contribution. 

For the vast majority of FRBs, which are faint, our precision is instead limited by the additional uncertainty incurred in marginalizing over the differential ionospheric contribution, making these subradian systematics a higher-order concern; presently, we account for them only in the final error budgeting as by convolving the likelihood with a baseline-dependent Gaussian systematic floor determined ad-hoc based on the test localizations presented in Section \ref{sec:final}. We defer a full treatment of these subradian systematics to a future analysis with our tracking-beam \citep{tracking_beams} upgraded system, which will allow calibrators to be observed for far longer than the current full-array captures, so that the calibration is never limited by calibrator signal-to-noise.

\section{Global Calibration}
\label{sec:global_cal}
The per-burst fit of Section~\ref{sec:visibility_model} determines each FRB's position together with a handful of per-source nuisance parameters, but is globally dependent on a set of ``meta-parameters''---the Outrigger station geometry and the form of the prior on the differential ionosphere---that are shared across the survey and fixed once from calibrator observations, rather than re-fit for each burst. Here we describe how these global quantities are calibrated and validated.

\subsection{Datasets}
\label{sec:calibration_dataset}
\begin{figure}
    \centering
    \includegraphics[width= \linewidth]{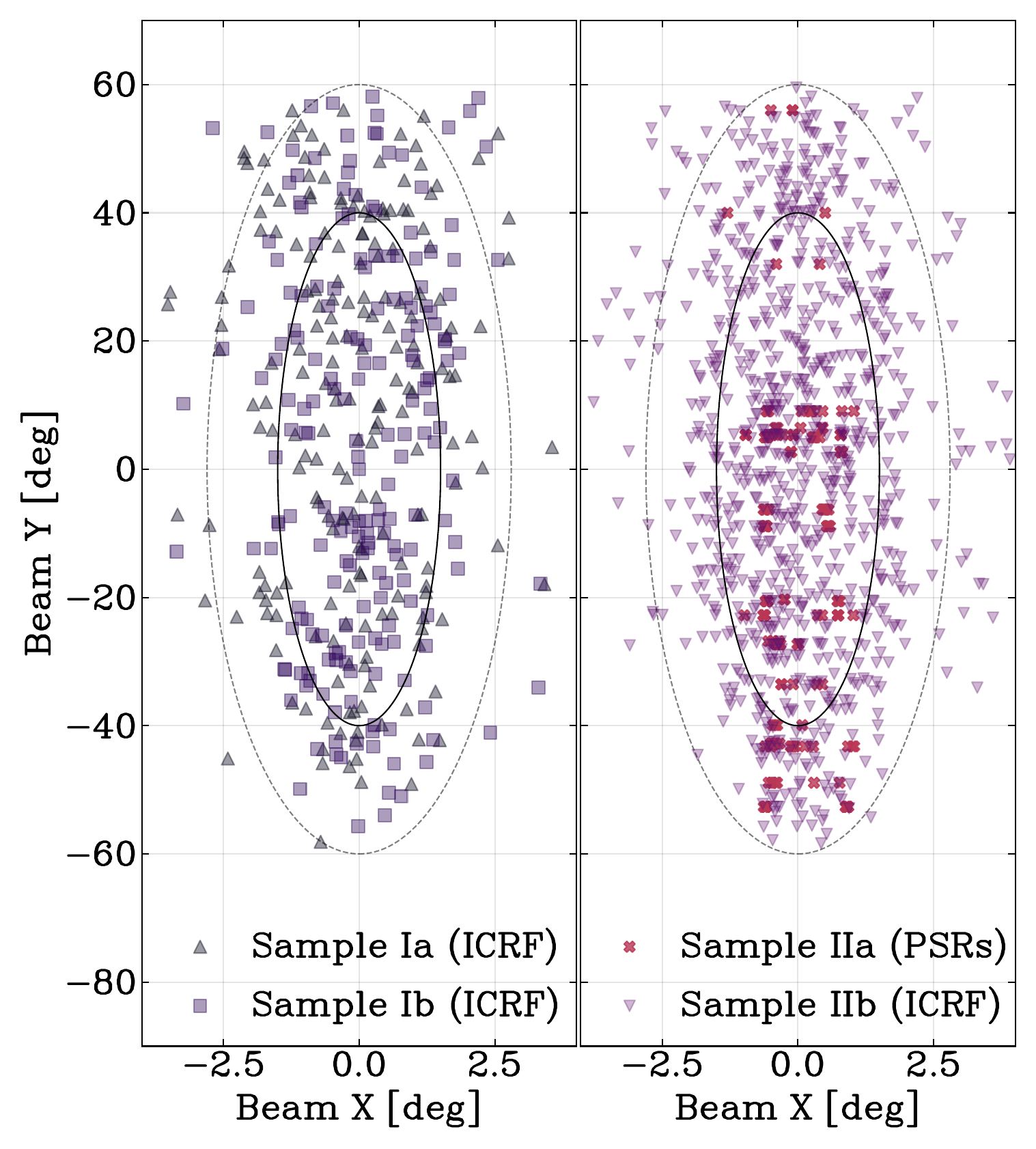}
    \caption{Distribution of all test localization sources used throughout this work on the sky in topocentric beam coordinates relative to CHIME zenith. ``Sample I'' (left) consists of the brightest in-beam RFC calibrators \citep{petrov2021wide} observed contemporaneously with FRB baseband acquisitions, and ``Sample II'' (right) consists of both pulsar observations with known VLBI positions (IIa) and RFC sources (IIb) observed contemporaneously with FRB baseband acquisitions.}
    \label{fig:test_sample_dist}
\end{figure}

First, we define the dataset used to characterize our systematics, inform our priors, and validate the fringe-fit model of Section~\ref{sec:visibility_model}. 
To reflect typical integration times of observed FRBs and calibrator S/Ns, as well as
the effect of the dispersive sweep on the ionospheric fringe fitting, we use in-beam calibrators within \ncaldump FRB baseband dumps to ``globally'' calibrate our localizations.
The typical distribution of in-beam calibrators observed contemporaneously with a full-array baseband acquisition is shown in Figure~\ref{fig:calibrator_FOV}, with $\gtrsim 5$ within the full $\sim 200\,\mathrm{deg}^2$ field of view on all baselines.  The distribution of all sources used in this sample are shown in Figure \ref{fig:test_sample_dist}, which span a broad range of sources that cover the full primary beam.  We divide our dataset into two broad samples: ``Sample I'' consists entirely of the brightest ~$\sim$ 200 in-beam calibrators from the Radio Fundamental Catalog (RFC) \citep{petrov2021wide} observed contemporaneously with FRB baseband acquisitions from Feb 2025-Feb 2026, randomly divided into Sample Ia used for the baseline offset and thin shell prior width ``meta-fits" performed in Sections \ref{sec:baseline_offset} and \ref{sec:thin_shell} (Figures \ref{fig:baseline_offset} and \ref{fig:ionosphere_histogram}), and Sample Ib used as an independent test of the fit results (Figures \ref{fig:baseline_offset} and \ref{fig:ionosphere_prior_ll}). ``Sample II'' is used to perform the end-to-end test localizations in Section \ref{sec:final}, consisting of pulsar observations with known VLBI positions (IIa) and RFC sources (IIb) observed contemporaneously with FRB baseband acquisitions from Feb 2025-Feb 2026 to supplement the lack of declination coverage provided by the pulsars. 

To test how our delay residuals vary with S/N and bandwidth, we repeat the analysis with systematically degraded copies of our bright and broadband calibrators in Sections \ref{sec:thin_shell} and \ref{sec:final}. These data are constructed with reduced S/N by shortening the integration time, or with reduced bandwidth by flagging frequencies in the visibilities. 


\subsection{Baseline offset}
\label{sec:baseline_offset}

As done in \cite{lanman2024kko}, we take the CHIME position to be correct\footnote{The VLBI delay depends primarily on the baseline vector so an error in CHIME's absolute position is absorbed into the fitted Outrigger position and leaves the baseline---and hence the FRB localizations---unchanged; it enters only through second-order aberration terms. Even a $\sim 100$\,m absolute error only shifts the reference station's rotational velocity by $\Omega_\oplus\,\delta r\sim7\times10^{-3}$\,m\,s$^{-1}$, and hence the delay by $\sim(\delta V/c)\,\tau_{\rm geo}\sim0.2$\,ps, compared with $\sim\delta r/c\sim330$\,ns for the same error in the baseline vector.} and sample Ia (correlated to their true positions) to fit for the outrigger station position on a given baseline using the fringe fit of the form presented in Equation \ref{eq:fringe_model} with the delay contribution modeled by Equation \ref{eq:baseline_offset}. the degeneracy between the ionosphere and delay in the complete absence of an ionospheric prior can be disentangled at the $\sim$~ns level with bright, broadband observations. 

Our fit is shown in Figure \ref{fig:baseline_offset} and our results are cross-validated with sample Ib, where we find that the angular 
offset residuals are detrended to the $\sim$ns level after the fit, corresponding to a constraint on the telescope positions to $<$1~m. 

\subsection{Ionosphere}
\label{sec:thin_shell}
For many FRBs, the S/N and effective bandwidth of the cross-correlated visibilities are not large enough to break the degeneracy between the ionospheric and geometric delays to the precision our target spec requires without an ionospheric prior \citep[see also Appendix~A of][]{outriggers_overview}. However, because we typically observe five or more in-beam calibrators simultaneously with the target (Figure~\ref{fig:calibrator_FOV})---roughly one ionospheric sightline per $\sim 20^\circ$ in declination across the $\sim 200\,\mathrm{deg}^2$ field of view---we can independently determine the ionospheric column at each station and impose it as a prior when fringe-fitting the target.

\begin{figure*}[p]
    \centering
    \includegraphics[width= \textwidth]{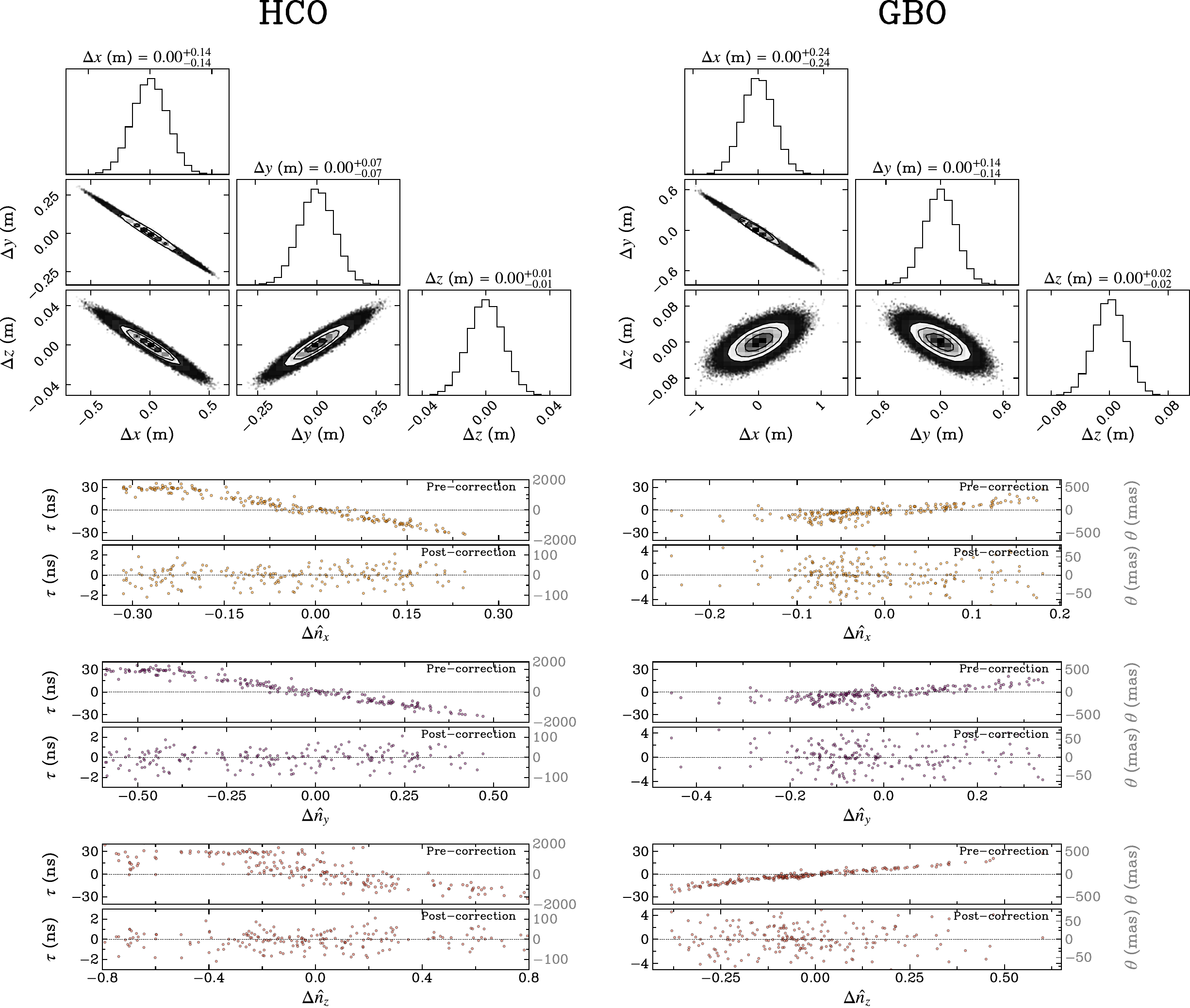}
    \caption{Top panels: best fit parameters for the station coordinates at HCO (left) and GBO (right) in geocentric coordinates to sample Ia described in Section \ref{sec:calibration_dataset}. Parameters have been re-centered about the best final fit value for each telescope, which are \HCOPOS for HCO and \GBOPOS for GBO. Bottom panels: Residual delays pre and post baseline offset correction for sample Ib on the CHIME-HCO (left) and CHIME-GBO (right) baselines as a function of projected
    calibrator-target separation onto different components of the baseline vector in geocentric coordinates.}
    \label{fig:baseline_offset}
\end{figure*}

\begin{figure}[t]
    \centering
    \includegraphics[width= \linewidth]{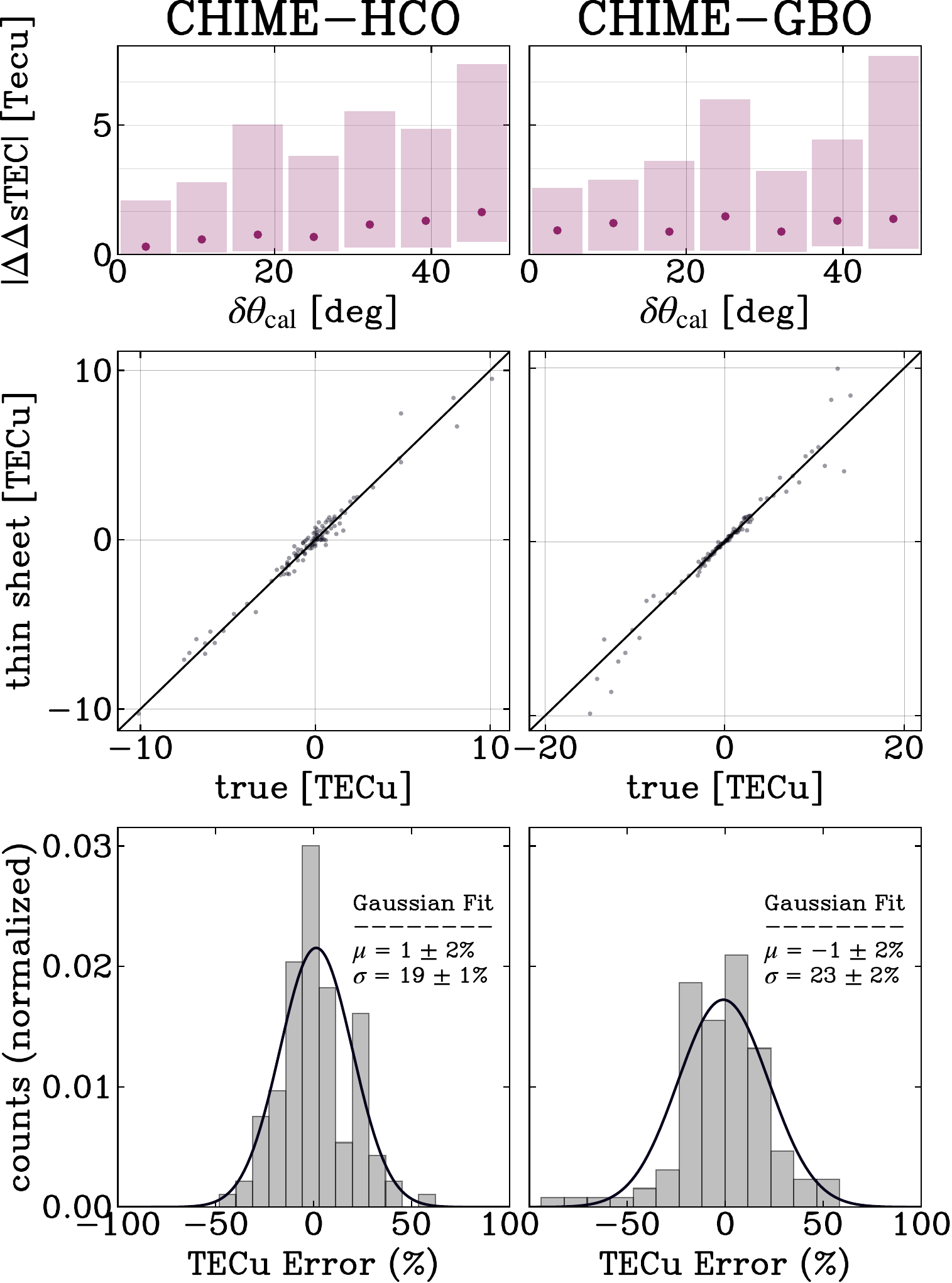}
    \caption{Top: Measured double-differential slant TEC as a function of angular separation for CHIME-HCO (left) and CHIME-GBO (right) on the sky for the high signal-to-noise test sample Ia, where the TEC solution is known to subnanosecond precision (as justified by delay residuals in Figure \ref{fig:baseline_offset}). The shaded regions show the 10th and 90th percentiles. The magnitude of the differential slant TEC even at the smallest angular separations seen by our currently accessible calibrator grid necessitates an ionospheric fringe fit to reach our astrometric spec. Middle: Comparison of model-fit ionosphere solution via the procedure described in Section \ref{sec:thin_shell} to the directly measured ionosphere solution. Bottom: the corresponding fractional error distribution justifies our choice of prior width.}
    \label{fig:ionosphere_histogram}
\end{figure}

For low-S/N targets we tighten the fit with a prior on the differential
sTEC constructed from a thin-shell model of the ionosphere~\citep[e.g.,][]{martin2016limits} fit to the
surrounding calibrators. We model the ionosphere as an infinitesimally
thin spherical shell concentric with the Earth, at fixed height $h$ above
the surface (Earth radius $R_\oplus$), and take the zenith
angle $\chi$ of the ionospheric pierce point, for a source observed at elevation $\varepsilon$ from a ground station, to be $ \sin \chi = \frac{R_\oplus}{R_\oplus + h}\,\cos\varepsilon$
\footnote{For our observations we adopt $R_\oplus = 6{,}378{,}137$\,m and
$h = 200$\,km}. We model the slant TEC (sTEC) for a station with vertical TEC (vTEC) $\mathcal{T}$ as
\begin{equation}\label{eq:stec}
    \mathrm{sTEC} = \frac{\mathcal{T}}{\cos\chi},
\end{equation}
so the double-differential slant TEC between a
target $T$ and a phase-reference calibrator $C$ observed from two
stations $a$ and $b$ is modeled as
\begin{equation}\label{eq:dd_exact}
\begin{split}
    \Delta\Delta\,\mathrm{sTEC}^{TC}_{ab}
    = &\mathcal{T}_a\!\left(\frac{1}{\cos\chi^T_a}
      - \frac{1}{\cos\chi^C_a}\right)\\
    - &\mathcal{T}_b\!\left(\frac{1}{\cos\chi^T_b}
      - \frac{1}{\cos\chi^C_b}\right).
\end{split}
\end{equation}
where $\mathcal{T}_a,\mathcal{T}_b$ are the per-station vertical TEC values we fit for using the in-beam calibrators. We validate this model---and quantify its prior width---by leave-one-out cross-validation: for each observation of $N$ calibrators, we hold one out as a mock ``target'', fit $\mathcal{T}_a,\mathcal{T}_b$ to the remaining $N-1$, and compare the predicted $\Delta\Delta\,\mathrm{sTEC}^{TC}_{ab}$ at the held-out position to its directly measured value (Figure~\ref{fig:ionosphere_histogram}). The held-out value can be taken as truth because these calibrators are bright enough that their delays meet our $\sim$50\,mas/$\sim$1\,ns spec (Figure~\ref{fig:baseline_offset}), so their ionospheric solutions are measured directly rather than inferred from the model. We find that the model can estimate the differential sTEC at the target location to $\sim$20$\%$ accuracy, which we adopt as the prior width floor (which is added in quadrature with the per-event statistical uncertainties on the inferred prior mean from the in-beam calibrator fits).

\begin{figure}
    \centering
\includegraphics[width= \linewidth]{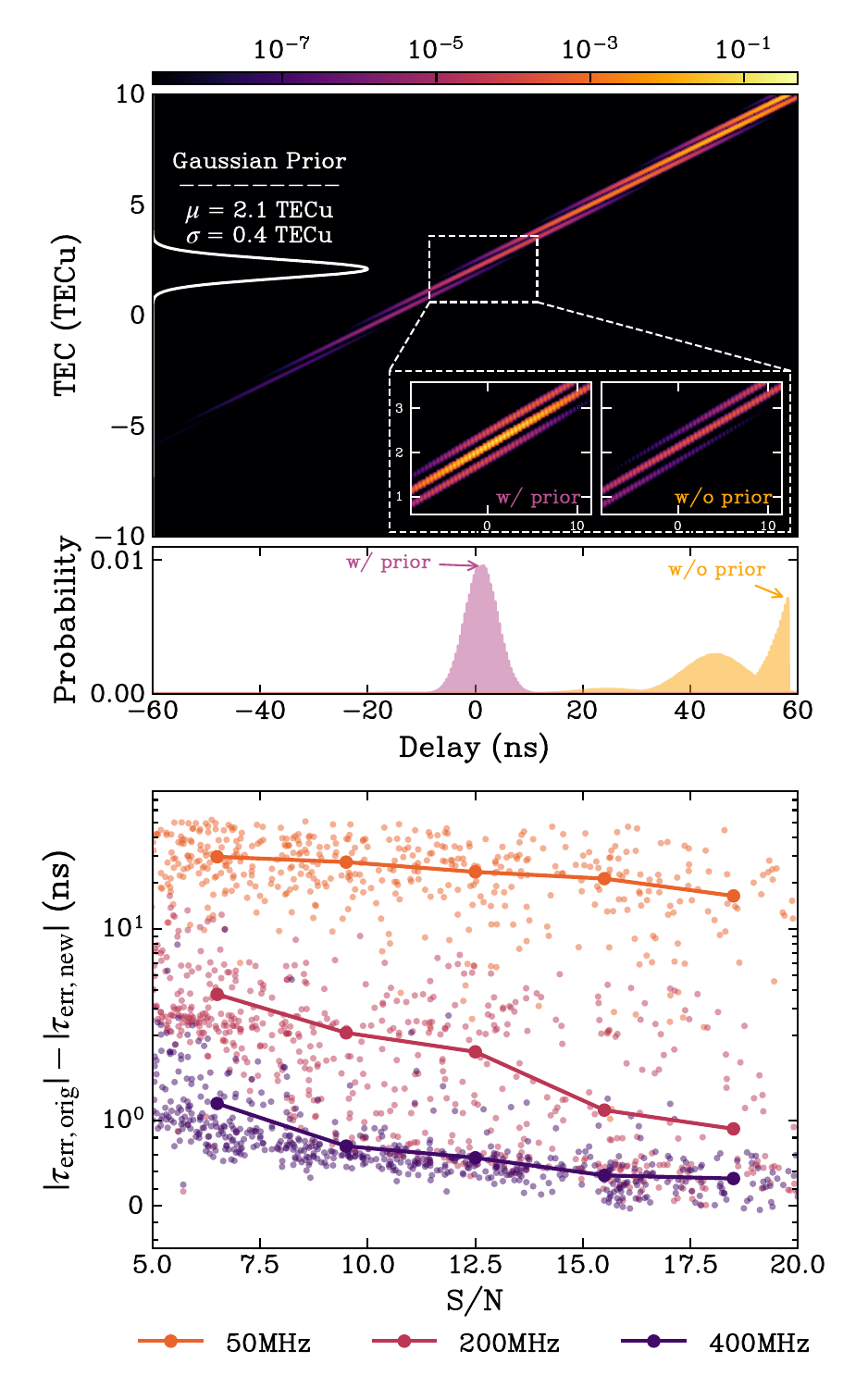}
    \caption{Fringe fit improvement with ionospheric prior derived from RFC sample Ib, with the target signal-to-noise degraded by \emph{decreasing} the integration time and the bandwidth degraded by flagging out subsections of the band in the cross correlated visibilities (see Section \ref{sec:global_cal} for full description). Top panel: example of a fringe fit posterior as a function of $\Delta \Delta$sTEC and delay for a single target, with posterior after marginalizing over $\Delta \Delta$sTEC shown underneath without the thin-shell prior (yellow) and with the prior (purple)  derived from simultaneous observations of in-beam calibrators. Bottom  panel: improvement in delay error (magnitude of old delay residual minus magnitude of new delay residual) for targets with 50MHz, 200MHz, and 400MHz bandwidths, with the overlaid curves showing the average improvement for different signal-to-noise bins.   Unsurprisingly, the largest delay improvements are seen at narrow bandwidths and low signal-to-noise.}
\label{fig:ionosphere_prior_ll}
\end{figure}

The resulting improvement in delay accuracy is shown in Figure \ref{fig:ionosphere_prior_ll}, where the astrometric residuals decrease most at narrow bandwidth and low S/N---the regime in which most of our FRBs are detected---both for individually S/N- and bandwidth-degraded calibrators and across the larger sample (and the test localizations of Section \ref{sec:final}).

\section{End-to-end test localizations}
\label{sec:final}
Armed with the results of the previous sections, we now perform a full end-to-end validation with over \npulsardumps  ``single epoch" 
localizations from 20 unique pulsars at four different bandwidths 
(full band, 200MHz, 100MHz, and 50MHz) where multiple bandwidths for a broadband source are obtained by flagging out subsections of the band. We supplement this pulsar sample (Sample IIa shown in Figure~\ref{fig:test_sample_dist}) with in-beam continuum sources observed simultaneously with FRBs in baseband acquisitions (Sample IIb shown in Figure~\ref{fig:test_sample_dist}) to account for any potential systematics arising from DM sweeps or beam position that are not reflected in the distribution of the pulsar sample, yielding $\gtrsim 1000$ test localizations total. We use astrometric measurements as measured by the VLBA for the pulsars B0136+57, B2154+40, B0450+55, B1541+09, B2053+36, and B2310+42 \citep{Chatterjee_2009ApJ}; B0329+54 \citep{Kumar_2025}; B1642$-$03, B0823+26, B0611+22, B2044+15, B1322+83, and B1917+00 \citep{Deller_2019}; B2021+51, B1929+10, and B2020+28 \citep{Kirsten_2015}; B0531+21 \citep{Crab_Lin}; B0355+54 \citep{Chatterjee_2004ApJ}; B0834+06 \citep{Liu_2016}; B2016+28 \citep{Brisken_2002ApJ}; and B0919+06 \citep{Fomalont_1999AJ}. We follow the 
localization procedure described in Section 2 with the calibration corrections 
applied from Section 3.

\begin{figure*}[p]
    \centering
    \includegraphics[width=\textwidth,height=0.9\textheight,keepaspectratio]{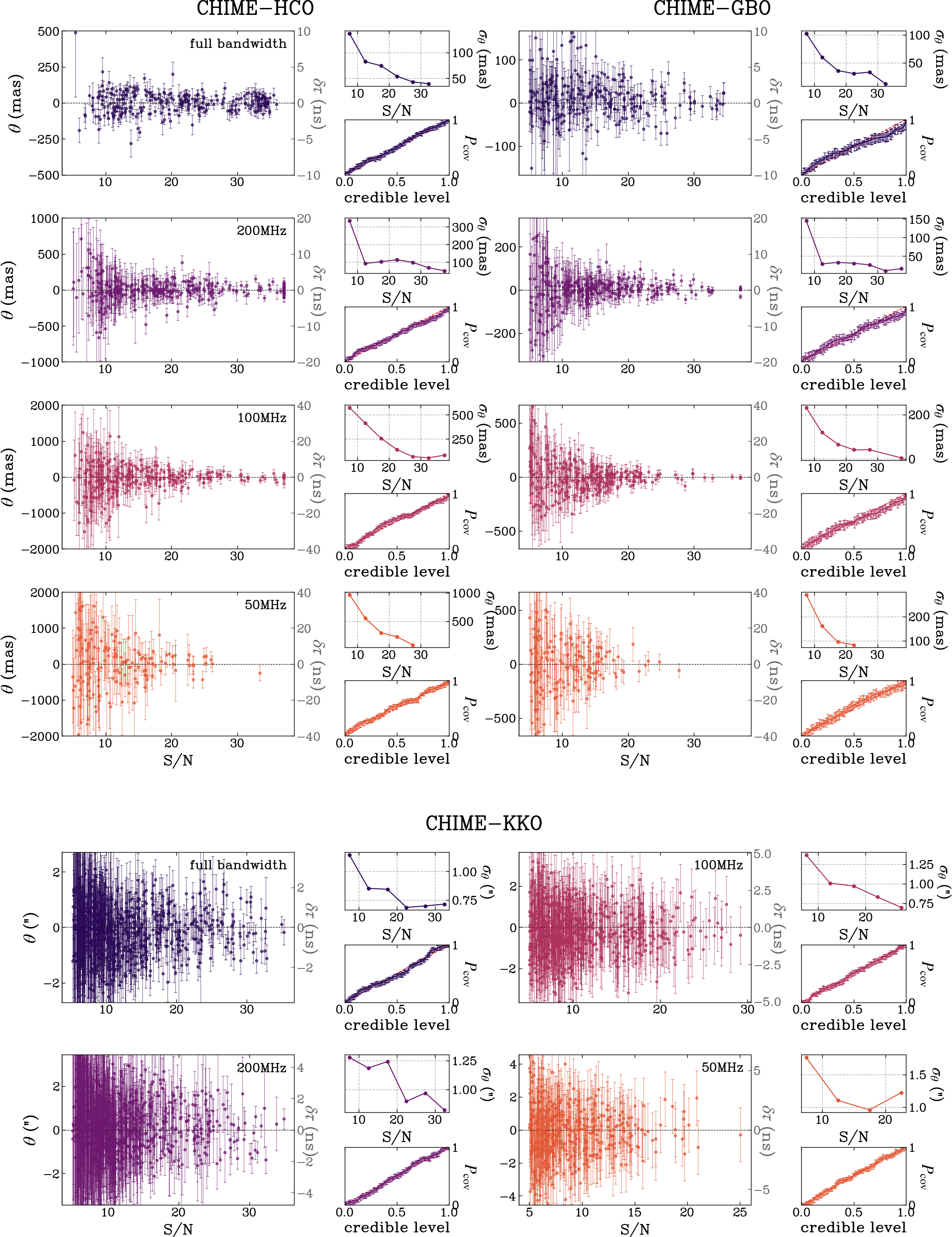}
    \caption{Distribution of angular offset errors from test localizations on the CHIME-HCO (left) and CHIME-GBO (right) 
    baselines, with uncertainties derived from the fringe fit posterior. To the right of each distribution, we show the average astrometric error
    in different signal-to-noise bins (top) with our signal to noise defined by the statistic presented in Appendix \ref{appendix:snr} Note that the errors are dependent on a large number of parameters (calibrator target separation, target distribution of signal-to-noise across the band, target-calibrator signal-to-noise overlap across the band, RFI, etc) beyond just S/N and bandwidth. Because the shape of the posterior is often not-gaussian, we perform a coverage test by showing the fraction of sources within a quoted credible level given by the fringe fit posterior as a function of the theoretical number of sources that should be within that level if the contours perfectly described the astrometric errors. Error bars on the coverage test are bootstrapped over 8 random subsamples of the data. We find our distribution to be consistent with the one-to-one line corresponding to accurate calibration.}
    \label{fig:final_test_loc}
\end{figure*}

The corresponding distribution of our errors is provided in Figure \ref{fig:final_test_loc}: uncertainties 
are derived directly from the shape of the fringe fit posterior rather than applied in an ad-hoc manner. We assess the accuracy of these statistical contours through a coverage test where we compute the fraction of sources whose correct localization position lies within a quoted credible level given by the fringe fit posterior as a function of the theoretical number of sources that should be within that level if the contours are correct; we find a strong one-to-one correspondence, indicating that the statistical
contours obtained from our current procedure well describe the true astrometric errors. 

The final precision of our combined localizations range from  $\sim 0.9"$ for the faintest, most narrowband bursts to $\sim$20mas for the brightest, broadband bursts, with a current systematics floor we attribute to an instrumental beam phase residuals. For bursts that are observed to be broadband in the cross-correlation, our localizations regularly exceed a precision of $\sim$50$\times$100 mas except for the faintest bursts, with particularly bright observations seeing a localization precision of $\sim$10$\times$30 mas. As expected, the astrometric precision is degraded for narrowband bursts where the degeneracy between the ionosphere and the geometric delay becomes more difficult to disentangle. However, we note that most narrowband FRBs detected by CHIME are observed to be repeaters \citep{Pleunis_2021}, which means that we can expect to obtain a refined position for a subset of narrowband localized bursts that are observed over multiple epochs–––this technique will be demonstrated in a future work \citep{hewitt2026}. Finally, we note that we find no statistically significant correlation between astrometric error and DM sweep or astrometric error and beam position at the values probed (shown in Figure \ref{fig:test_sample_dist}). 

The localization error for the majority of bursts is dominated by the ionospheric contribution, which contains a strong degenearcy with the geometric delay. This is reflected in the fact that that the \emph{delay} error tends to be the smallest on the shortest baseline (CHIME-KKO) where the differential ionosphere is smaller in conjunction with a denser calibrator grid tends provides a much tighter prior on the ionosphere. Although this is alleviated by the wide-field ionospheric modeling described in Section \ref{sec:thin_shell}, we are currently undergoing a campaign to build a much denser calibrator grid \citep{atkinson2026} observable with our future tracking beam system \citep{tracking_beams} that is expected to provide much tighter constraints on the ionosphere. 


\section{Discussion, Future Work, and Conclusion}
\label{sec:discussion}
In this work, we have presented the end-to-end procedure used to localize single-pulse sources with the fully comissioned CHIME/FRB Outriggers network. We validated this procedure on a sample of over \npulsardumps\ single-epoch continuum source and pulsar localizations evaluated across four bandwidths for a total of $\sim 1000$ test localizations. The statistical contours derived from the fringe-fit posterior faithfully describe the true astrometric errors, with final combined precisions for the full array ranging from $\sim 0.9''$ for the faintest, most narrowband bursts to $\sim 20$\,mas for the brightest, broadband bursts. CHIME/FRB Outriggers thus meets, and under favorable conditions exceeds, the $\sim 50$\,mas astrometric specification required for its central science goals.

We expect our astrometric performance to improve appreciably with the forthcoming tracking-beam system. By allowing calibrators to be observed for far longer than is possible with the current full-array captures, the tracking-beam system provides access to a much denser calibrator grid that can significantly reduce the ionospheric contribution to the overall error for FRBs that are faint and potentially enable the full treatment of the subradian systematics that we presently absorb into the systematic floor (Section \ref{sec:superresolve}). We therefore anticipate that the astrometric precision for the majority of FRBs to improve well beyond the current precision in future analyses.

More broadly, the astrometric techniques and performance demonstrated here on a large suite of test localizations establishes the technical feasibility of precision VLBI astrometry with a wide-field survey instrument. CHIME/FRB Outriggers is the first of a new generation of such instruments, and its success in routinely localizing one-off FRBs to $\sim$ 50 milliarcsecond precision \citep[e.g.][]{Blanchard_2025,RBFloat_2025} charts a path toward the large, environment-resolved samples needed to disentangle FRB progenitor channels at the population level. 

\section{Acknowledgments}
C. L. acknowledges support from the Miller Institute for Basic Research at UC Berkeley. K.W.M. is supported by an NSF Grant (2008031). D.M.H. and Z.P. are supported by an ERC Starting Grant (`EnviroFlash’; Grant agreement No. 101223057); and an NWO-Veni grant (VI.Veni.222.295). A.A. acknowledges the support
from the Natural Sciences and Engineering Research Council of
Canada (NSERC-CGSM). S.C. acknowledges support from the National Science Foundation (NSF AAG 2511105). J.M.P. acknowledges the support of an NSERC Discovery Grant (RGPIN-2023-05373). K.N. acknowledges support by NASA through the NASA Hubble Fellowship grant \# HST-HF2-51582.001-A awarded by the Space Telescope Science Institute, which is operated by the Association of Universities for Research in Astronomy, Incorporated, under NASA contract NAS5-26555. A.B.P.~acknowledges support by NASA through the NASA Hubble Fellowship grant \mbox{HST-HF2-51584.001-A} awarded by the Space Telescope Science Institute, which is operated by the Association of Universities for Research in Astronomy, Inc., under NASA contract \mbox{NAS5-26555}. A.B.P.~also acknowledges prior support from a Banting Fellowship, a McGill Space Institute~(MSI) Fellowship, and a Fonds de Recherche du \mbox{Qu\'ebec -- Nature} et Technologies~(FRQNT) Postdoctoral Fellowship. V.S. is supported by a Fonds de Recherche du Quebec—Nature et Technologies (FRQNT) Doctoral Research Award.

We acknowledge that CHIME is located on the traditional, ancestral, and unceded territory of the Syilx/Okanagan people. We are grateful to the staff of the Dominion Radio
Astrophysical Observatory, which is
operated by the National
Research Council Canada. 
CHIME is funded by a grant from the Canada Foundation 
for Innovation (CFI) 2012 Leading Edge Fund (Project 31170) 
and by contributions from the provinces of British Columbia, 
Qu\'ebec and Ontario. The CHIME/FRB Project is funded by a 
grant from the CFI 2015 Innovation Fund (Project 33213) and 
by contributions from the provinces of British Columbia and 
Qu\'ebec, and by the Dunlap Institute for Astronomy and 
Astrophysics at the University of Toronto. 
Additional support was provided by the Canadian 
Institute for Advanced Research (CIFAR), McGill 
University and the McGill Space Institute via the 
Trottier Family Foundation, and the University of 
British Columbia. 
The CHIME/FRB Outriggers program is funded by 
the Gordon and Betty Moore Foundation
and by a National Science Foundation (NSF) grant (2008031).
\appendix
\section{Detection significance}
\label{appendix:snr}
Because the Outriggers contain a smaller collecting area than CHIME, not every burst detected by the central core will be observable in VLBI. We therefore require a carefully-defined detection significance to robustly assess whether a given fit can be taken as a reliable measurement. To define our statistic, we start with Wilks' theorem, which states that the log-likelihood ratio between the best fit model ($\mathcal{L}_\mathrm{best}$) and the null ($\mathcal{L}_0$ with $S_\nu = 0$) follows a $\chi^2$ distribution with $d$ degrees of freedom equal to the number of free parameters:
\begin{equation}
    \label{eq:chi_vals}
    \Delta\ln\mathcal{L}
    = 2\left(\ln\mathcal{L}_\mathrm{best} - \ln\mathcal{L}_0\right) \sim \chi^2_d.
\end{equation}
In practice, however, the number of degrees of freedom given nominally by Wilks' theorem often does not describe the null distribution when the null lies on the boundary of the domain for a parameter (which applies to our case where $S_\nu = 0$). Rather than assume the nominal value and apply an explicit trials correction, we measure a distribution of the statistic given by Equation \ref{eq:chi_vals} using  off-lags (delays far from any fringe) where the visibilities contain only noise and fit for an effective $d$. We then take the $p$-value directly from the $\chi^2$ survival function, $p = 1 - F_{\chi^2}(\Delta\ln\mathcal{L},\, d)$, and take the equivalent Gaussian
significance to be 
\begin{equation}\label{eq:sigma}
    \sigma = \Phi^{-1}(1 - p),
\end{equation}
where $\Phi^{-1}$ is the inverse of the standard normal cumulative distribution function.

We assess the validity of this statistic using visibilities gated on an off-pulse time window so that no signal is present, but otherwise processed exactly as a real burst (Section \ref{sec:final}). We compare the reported false-positive probability (Equation \ref{eq:sigma}) to the fraction of nulls that actually exceed it. We find a one-to-one correspondence (e.g. $5\%$ of these pure-noise fits in fact reach a significance that corresponds to a 5\% false-positive probability) which indicates our significances are well calibrated. We also note that, for observations we know contain a real signal (the ``on-pulse'' sample of Section \ref{sec:final}), we find that the likelihood remains calibrated up to the detection threshold, and below a S/N of $\sim 5$ the localizations widen to a flat probability distribution (an uninformative localization).

\section{Amplitude-marginalized likelihood}
\subsection{Single calibrator}
We can express the general form of the likelihood provided by Equation \ref{eq:general_loglik} for a single frequency $\nu$ and polarization $\alpha$ in the form
\begin{equation}\label{eq:loglik_single}
    \ln\mathcal{L}_{\alpha\nu} = -\frac{|V_{\alpha\nu} - s_\alpha S_\nu P_\nu|^{2}}{2\sigma_\nu^{2}}
    = -\frac{|V_{\alpha\nu}|^{2}}{2\sigma_\nu^{2}} + \frac{s_\alpha R_{\alpha\nu}}{\sigma_\nu^{2}}\,S_\nu - \frac{s_\alpha^{2}}{2\sigma_\nu^{2}}\,S_\nu^{2},
\end{equation}
where $R_{\alpha\nu}\equiv\mathrm{Re}[V_{\alpha\nu}\bar P_\nu]$. Adding the log prior $\ln p(S_\nu) = -(S_\nu-\bar S_\nu)^{2}/2\delta S_\nu^{2}$ and marginalizing provides
\begin{equation}
    \mathcal{L}_{{\alpha\nu}, \mathrm{marg}} = e^{-|V_{\alpha\nu}|^{2}/2\sigma_\nu^{2}}\int_{0}^{\infty} e^{f(S_\nu)}\,dS_\nu = L_{0,\nu} \int_{0}^{\infty} e^{f(S_\nu)}\,dS_\nu
\end{equation}
where $L_{0,\nu}$ is the likelihood  of the null used in our Wilks statistic (Eq.~\ref{eq:chi_vals}) and 
\begin{align}\label{eq:f_single}
    f(S_\nu) &= \frac{s_\alpha R_{\alpha\nu}}{\sigma_\nu^{2}}\,S_\nu - \frac{s_\alpha^{2}}{2\sigma_\nu^{2}}\,S_\nu^{2} - \frac{(S_\nu-\bar S_\nu)^{2}}{2\delta S_\nu^{2}} \\
    &= -\tfrac12\Lambda_{\alpha\nu}(S_\nu-\mu_{\alpha\nu})^{2} + \tfrac12\Lambda_{\alpha\nu}\mu_{\alpha\nu}^{2} - \bar S_\nu^{2}/2\delta S_\nu^{2}
\end{align}
with
\begin{align}\label{eq:lam_mu_single}
    \Lambda_{\alpha\nu} &\equiv \frac{s_\alpha^{2}}{\sigma_\nu^{2}} + \frac{1}{\delta S_\nu^{2}}  \\
    \mu_{\alpha\nu} &\equiv \frac{s_\alpha R_{\alpha\nu}/\sigma_\nu^{2} + \bar S_\nu/\delta S_\nu^{2}}{\Lambda_{\alpha\nu}}
\end{align}
The Gaussian integral over $S_\nu\ge0$ then gives
\begin{equation}\label{eq:halfline}
    \int_{0}^{\infty} e^{f(S_\nu)}\,dS_\nu
    = \sqrt{\frac{2\pi}{\Lambda_{\alpha\nu}}}\;\tfrac{1}{2}\,\mathrm{erfc}\!\left(-\frac{z_{\alpha\nu}}{\sqrt 2}\right)\,
    \exp\!\left(\tfrac{1}{2}\Lambda_{\alpha\nu}\mu_{\alpha\nu}^{2} - \frac{\bar S_\nu^{2}}{2\,\delta S_\nu^{2}}\right),
\end{equation}
with
\begin{equation}
z_{\alpha\nu}\equiv\sqrt{\Lambda_{\alpha\nu}}\,\mu_{\alpha\nu}
\end{equation}
which after summing over the statistically independent channels and polarizations is equivalent to the exponential of the amplitude-marginalized log likelihood for a single calibration solution provided in Equation ~\ref{eq:loglike_marg}.
\subsection{Multiple calibrators}
\label{appendix:joint_cal_lik_marg}
Packing the phase model for each calibrator $P_c = \exp[2\pi i(\nu\tau + \kdm\,\mathrm{TEC}_c/\nu)]$ into a vector $\mathbf{p}$ and our visibilities $V_{\alpha,\nu}$ into a vector $\mathbf{y}$, we have the model vector $\mathbf{m} = s_\alpha S_\nu\,\mathbf{p}$. The vector form of our likelihood can be expressed as
\begin{equation}\label{eq:quadratic_expand}
\ln\mathcal{L}_{\alpha\nu}  = -\tfrac12(\mathbf{y}-\mathbf{m})^{\dagger}C^{-1}(\mathbf{y}-\mathbf{m})
    = -\tfrac12\,\mathbf{y}^{\dagger}C^{-1}\mathbf{y} + s_\alpha S_\nu\,\mathrm{Re}\!\left[\mathbf{p}^{\dagger}C^{-1}\mathbf{y}\right]
    - \tfrac12\,s_\alpha^{2}S_\nu^{2}\,\mathbf{p}^{\dagger}C^{-1}\mathbf{p},
\end{equation}
where the latter equivalence holds because $C^{-1}$ is Hermitian ($\mathbf{p}^{\dagger}C^{-1}\mathbf{p}$ is real). Note that if we define
\begin{align}
    U_{\alpha\nu} &\equiv \mathbf{p}^{\dagger} C^{-1}\mathbf{p}, &
    W_{\alpha\nu} &\equiv \mathrm{Re}\!\left[\mathbf{p}^{\dagger} C^{-1}\mathbf{y}\right],
\end{align}
this can be re-expressed as 
\begin{align}
    \ln\mathcal{L}_{\alpha\nu}  &= \ln \mathcal{L}_0+ s_\alpha S_\nu\,\!\left[W_{\alpha\nu}\right]
    - \tfrac12\,s_\alpha^{2}S_\nu^{2}\,U_{\alpha\nu}
\end{align}
Since $U_{\alpha\nu}$ and $ W_{\alpha\nu}$ are independent of $S_\nu$, this provides an identical expression to that given by Equation \ref{eq:loglik_single} with the substitutions
\begin{align}\label{eq:UW_def}
    \frac{R_{\alpha\nu}}{\sigma_\nu^{2}} &\longrightarrow\ W_{\alpha\nu}\equiv\mathrm{Re}\!\left[\mathbf{p}^{\dagger}C^{-1}\mathbf{y}\right], & \frac{1}{\sigma_\nu^{2}} &\longrightarrow\ U_{\alpha\nu}\equiv\mathbf{p}^{\dagger}C^{-1}\mathbf{p}
\end{align}
and the subsequent joint amplitude-marginalized likelihood therefore takes the analogous form
\begin{equation}\label{eq:joint_loglik_marg}
    \ln\mathcal{L}_{\mathrm{marg}} \propto
    \sum_{\alpha,\nu}\left[
    \tfrac{1}{2}\Lambda_{\alpha\nu}\,\mu_{\alpha\nu}^{2}
    - \frac{\bar S_\nu^{2}}{2\,\delta S_\nu^{2}}
    - \tfrac{1}{2}\ln\Lambda_{\alpha\nu}
    + \ln\!\left(\tfrac{1}{2}\,\mathrm{erfc}\!\left(-\frac{z_{\alpha\nu}}{\sqrt 2}\right)\right)
    \right],
\end{equation} 
with 
\begin{align}
    \Lambda_{\alpha\nu} &= s_\alpha^{2}\,U_{\alpha\nu} + \frac{1}{\delta S_\nu^{2}}, &
    \mu_{\alpha\nu} &= \frac{s_\alpha W_{\alpha\nu} + \bar S_\nu/\delta S_\nu^{2}}{\Lambda_{\alpha\nu}}, &
    z_{\alpha\nu} &= \sqrt{\Lambda_{\alpha\nu}}\;\mu_{\alpha\nu}.
\end{align}
\label{eq:joint_loglik_marg}

\onecolumngrid

\bibliography{references}{}

\bibliographystyle{aasjournal}
\end{document}